# Building Earth with pebbles made of chondritic components

Susmita Garai*, Peter L. Olson, Zachary D. Sharp
Earth and Planetary Sciences
University of New Mexico

**Abstract**

Pebble accretion provides new insights into Earth's building blocks and early protoplanetary disk conditions. Here, we show that mixtures of *chondritic components*: metal grains, chondrules, calcium-aluminum-rich inclusions (CAIs), and amoeboid olivine aggregates (AOAs) match Earth's major element composition (Fe, Ni, Si, Mg, Ca, Al, O) within uncertainties, whereas no combination of chondrites and iron meteorites does. Our best fits also match the $\varepsilon^{54}$Cr and $\varepsilon^{50}$Ti values of Earth precisely, whereas the best fits for chondrites, or components with a high proportion of E chondrules, fail to match Earth. In contrast to some previous studies, our best-fitting component mixture is predominantly carbonaceous, rather than enstatite chondrules. It also includes 15 wt% of early-formed refractory inclusions (CAIs + AOAs), which is similar to that found in some C chondrites (CO, CV, CK), but notably higher than NC chondrites. High abundances of refractory materials are lacking in NC chondrites, because they formed after the majority of refractory grains were either drawn into the Sun or incorporated into terrestrial protoplanets via pebble accretion. We show that combinations of Stokes numbers of chondritic components build 0.35-0.7 Earth masses in 2 My in the Hill regime accretion, for a typical pebble column density of 1.2 kg/m$^2$ at 1 au. However, a larger or smaller column density leads to super-Earth or moon-mass bodies, respectively. Our calculations also demonstrate that a few My of pebble accretion with these components yields a total protoplanet mass inside 1 au exceeding the combined masses of Earth, Moon, Venus, and Mercury. Accordingly, we conclude that pebble accretion is a viable mechanism to build Earth and its major element composition from primitive chondritic components within the solar nebula lifetime.

* Corresponding author: Susmita Garai *sgarai@unm.edu*

# 1. INTRODUCTION

Over the last several decades, impact-based accretion has become a well-established mechanism for planet formation (e.g., Safronov, 1972; Wetherill, 1980; Beaugé and Aarseth, 1990). According to this mechanism, planets grow over tens of million years via the accumulation of planetesimals and larger embryos (e.g., Chambers, 2004; Raymond et al., 2009; Morbidelli et al., 2012). Recently, an alternative mechanism called *pebble accretion* has emerged, in which a protoplanet grows by accretion of sub-millimeter to centimeter-size solids in just a few million years (Ormel and Klahr, 2010; Ormel, 2017; Johansen and Lambrechts, 2017) Pebble accretion has been successful in linking Jovian planets' masses and solar-like atmospheric metallicities, which requires growth within the protoplanetary disk lifetime of a few million years (Lambrechts and Johansen, 2012, 2014).

In addition to successful pebble models for Jovian planets, there have been a limited number of studies on pebble accretion for terrestrial planets. Recent simulations show that this rapid formation scenario has the potential to reconstruct the inner solar system including Earth (Johansen et al., 2015, 2021; Levison et al., 2015). Rapid pebble-driven growth has additional implications for terrestrial planets, including volatile enrichment by nebular ingassing (Sharp, 2017), partitioning of volatiles (Johansen et al., 2023b), Earth's silicate-metal segregation (Olson et al., 2022; Johansen et al., 2023a) and Hf-W isotopic system (Johansen et al., 2023a; Olson and Sharp, 2023), Earth's and Venus's atmospheric Ar and Ne compositions (Lammer et al., 2020), and Earth's overall oxidation state (Sharp, 2017).

The asteroid belt between Mars and Jupiter retains evidence of two chemically distinct regions: volatile-poor inner planets and volatile-rich outer planets (Warren, 2011). Chondritic meteorites derived from beyond Jupiter are carbonaceous or C, while those inward of Jupiter are non-carbonaceous or NC. C and NC chondrites also differ in their nucleosynthetic isotope compositions, suggesting incomplete mixing and partially different source materials (Kleine et al., 2020). The inner planets carry both C and NC chondritic signatures. For example, Earth has a nitrogen isotope composition (Sharp and Olson, 2022) and lithophile and refractory siderophile isotope contents (Javoy et al., 2010; Warren, 2011; Burkhardt et al., 2021) that suggest an NC source, whereas Si (Onyett et al., 2023) and Zn (Steller et al., 2022; Martins et al., 2023) isotopes indicate a large C component. As inferred from isotope studies on Ti, Zr, and Mo, a maximum of 4% (by mass) of C chondrite is predicted for Earth and Mars via an impact-based growth from a chondrite feedstock (Burkhardt et al., 2021). In contrast, Earth's Ca isotope compositions require a 40% contribution from C chondrites, which is more consistent with pebble-driven growth (Schiller et al., 2018). Similarly, pebble accretion successfully links the nucleosynthetic Fe isotope composition of Earth with CI chondrite (Schiller et al., 2020), although this interpretation has been questioned by Hopp et al. (2022) and Alexander (2022). Moreover, Earth's atmospheric noble gas composition resembles the C materials (Marty, 2022). Collectively, we do not yet understand the relative contributions of C and NC materials to Earth and their accretion mechanism.

Pebble accretion works as follows. In the absence of nebular gas, solids in a protoplanetary disk tend to follow Keplerian orbits, with a balance between gravity and rotational forces. If a gaseous nebula is present, then the pressure-sensitive gas molecules follow sub-Keplerian orbits with reduced velocities (Weidenschilling, 1977). Pebbles are therefore subject to gas drag from a headwind, causing loss of their angular momentum. As a consequence, the pebbles will drift radially toward the central star (Ormel and Klahr, 2010). In this way, pebbles can deliver outer solar system material to the inner solar system. The inward-drifting pebbles can then be captured gravitationally by growing planetesimals and planetary embryos, leading to a rapidly formed planet.

In this study, we explore some consequences of Earth's formation by pebble accretion. There exist ample motivations for this investigation. Young protoplanetary disks are huge reservoirs of pebbles (Testi et al., 2003; Andrews, 2015). Although the modern solar system is depleted in pebble-like solids, there is abundant evidence of pre-existing pebbles. In particular, chondrites are made of *components* such as metal grains, chondrules, and refractory inclusions, all of which have the dimensions appropriate for pebble accretion (Gooding, 1983; Kuebler et al., 1999; Scott and Krot, 2003; Teitler et al., 2010; MacPherson, 2014; Friedrich et al., 2015; Simon et al., 2018). These components must predate their parent chondrite bodies, meaning that they were isolated objects prior to accretion, and therefore qualify as pebbles. Recently, it has been proposed that the most abundant component in many chondrites – millimeter-size chondrules – were the building blocks for planetary accretion (Hewins and Herzberg, 1996; Johansen et al., 2015; Hartlep and Cuzzi, 2020). Hewins and Herzberg (1996) showed a chondrule-mantle composition similarity for Mg and Al, suggesting chondrules, rather than chondrites, as the source material to Earth. Also, it has been shown that Earth's major element composition can be approximated by a mixture of such chondritic components (Anders, 1977; Alexander, 2022). Therefore, pebble accretion with chondritic components as the pebble feedstock, is a good candidate for terrestrial planetary accretion.

We address two major unresolved issues regarding pebble accretion. First, we evaluate the efficiency and timing of accretion for growing an Earth-sized body with pebbles consisting of chondritic components. We apply a pebble accretion model for Earth constrained by the solar (Fe+Ni)/Si, in which C and NC chondrules, calcium-aluminum-rich inclusions (CAIs), amoeboid olivine aggregates (AOAs), and metal grains comprise the pebbles. We then explore the feasibility of making Earth's major element composition (Fe, Ni, Si, Mg, Ca, Al, O) by mixing variable proportions of 1) different chondrite and iron meteorite types, the so-called Chondrite model and 2) pebble components mentioned above, the so-called Component model. In the Chondrite model, we use fragments of Carbonaceous (C), Enstatite (E), and Ordinary (O) chondrites, plus iron meteorites as the pebble source. In the Component model, pebbles consist of metal grains, C, E, and O chondrules, CAIs, and AOAs.

The organization is as follows. In Section 2 we outline pebble accretion methodology, followed by Section 3 where we present our pebble accretion results. In Section 4 we describe our inversion methods for determining the bulk composition of Earth. Section 5 contains the inversion results. Comparisons between present and previous studies are given in Section 6. Lastly, we discuss our results in Section 7 and summarize our major findings in Section 8.

## 2. PEBBLE ACCRETION METHODOLOGY

Important variables and parameters are defined in Tables 2 and 3.

### 2.1. Pebble size

We define *pebbles* in pebble accretion as aerodynamically small particles with dimensions ranging from a fraction of a millimeter to a few centimeters that are weakly coupled to the gas (Ormel, 2017). Traditionally, pebble size is measured in terms of the Stokes number $St$, a dimensionless parameter equal to the product of orbital frequency $\Omega$ and friction time $\tau_f$ (Ormel and Klahr, 2010; Lambrechts and Johansen, 2012). In terms of pebble and disk properties (Johansen and Lambrechts, 2017),

$$St = \Omega \tau_f = \frac{\sqrt{2\pi} r_p \rho_p}{\Sigma_g}, \qquad (1)$$

where $r_p$ and $\rho_p$ are the radius and density of spherical pebbles and $\Sigma_g$ is the gas column density, the mass of gas in a cross section of unit area in the disk.

Rapid growth of a protoplanet by pebble accretion is most efficient with sub-mm to cm-size pebbles. Stokes numbers for such pebbles are typically between $10^{-3}$ and 1. In contrast, larger objects for which $St \gg 1$ are essentially uncoupled to the gas, whereas dust or µm size particles corresponding to $St \ll 10^{-3}$ are tightly coupled to the gas and therefore are not abundantly drawn into a growing planet. Accordingly, both larger and smaller objects are subject to slower and less efficient accretion in comparison to intermediate size pebbles.

**2.1.1. Stokes numbers of chondrites, iron meteorites, and chondritic components**

In our study, we consider chondritic components, fragments of chondrites, and iron meteorites as pebbles. It is worth noting two things here. First, chondrites and iron meteorites greatly exceeding 1 m in size qualify as gravitationally large objects ($St \gg 1$), and therefore do not satisfy the definition of pebbles. However, their fragments, if sufficiently small to have $St$ less than one, do fit the definition of pebbles.

Second, pebbles consisting of different chondritic components have different $St$ due to different densities and sizes. Table 1 shows that chondritic components vary between sub-mm to cm size, except matrix material, which is composed of dust grains (µm size). Based on their mean densities and sizes, we find that the $St$ of chondrules, refractory inclusions (AOAs and CAIs), and metal grains vary between $10^{-3}$ and $10^{-2}$ approximately, for a typical gas column density of $\Sigma_g = 10^3$ kg/m², corresponding to ~0.4 µbar at 1 au at the midplane of a standard disk (Johansen and Lambrechts, 2017). In contrast, the matrix material corresponds to $St < 10^{-5}$.

Metal grains may grow to form larger aggregates more efficiently than silicate pebbles due to magnetic fields, leading to chains of magnetized metallic grains (Kruss and Wurm, 2018). To account for this possibility, we also calculate the representative $St$ of metal aggregates, with an upper limit defined by their bouncing barrier, which we take to be 1 g (Zsom et al., 2010). As shown in Table 1, the $St$ for metal aggregates is between $10^{-2}$ and $10^{-1}$.

Since the growth of a protoplanet is controlled by the sizes of pebbles through their Stokes numbers, it is worth comparing how a protoplanet captures different chondritic components with different $St$. Following Olson et al. (2022), we calculated the pebble trajectories for $St = 10^{-1}$, $10^{-3}$, and $10^{-5}$ as shown in Figure S1 in the Supplementary Material. It is evident that for $St = 10^{-1}$, the pebble capture cross section is nearly twice the Hill radius $r_{Hill}$, which is defined by equation (8). On the other hand, $St = 10^{-3}$ leads to pebble capture over a narrower cross section of 0.4 $r_{Hill}$. Furthermore, a pebble with $St < 10^{-5}$ will only be captured over a narrower cross section of <0.06 $r_{Hill}$. Based on this comparison, we argue that chondrules, metal grains, and refractory inclusions are preferentially captured compared to the matrix among all the chondritic components. For the above reason, we consider a $St$ range between $10^{-3}$ and $10^{-1}$, thereby excluding matrix from our Component model.

## 2.2. Pebble accretion on Earth in the Hill regime

In our accretion calculations, we recognize two idealized groups of pebbles -- metal and silicate. Metal pebbles include the metal grains found as isolated grains in chondrites plus aggregates of these grains. Chondrules and refractory inclusions are considered to be silicate pebbles.

### 2.2.1. Compositional constraints from solar nebula

We constrain the nebular mass fractions of metal pebbles $x_m$ and of silicate pebbles $x_s$ required for an Earth-like composition, using the solar value of (Fe+Ni)/Si=1.84, which implies

$$x_m + x_s = 1; \frac{x_m + x_s X_i C_i^{\text{FeNi}}}{x_s X_i C_i^{\text{Si}}} = 1.84. \qquad (2)$$

Here, $X_i$ represents the fraction of $i^{th}$ silicate pebble type in the silicate pebble group and the summation is implied over the repeated index. $C_i^{\text{FeNi}}$ and $C_i^{\text{Si}}$ represent concentrations of Fe+Ni and Si, respectively, in the $i^{th}$ silicate pebble type. Values for $X_i$ in equation (2) is needed to calculate the nebular fraction and partial column density of each pebble group. We provide the procedure to find the values of $X_i$ in Section 5. Using the above expressions, the partial column densities of both pebble groups in the solar nebula can be written as

$$\Sigma_p^m = x_m \Sigma_p; \Sigma_p^s = x_s \Sigma_p, \qquad (3)$$

where $\Sigma_p^m$ and $\Sigma_p^s$ are the partial column densities of metal and silicate pebble groups, and $\Sigma_p$ is the total column density of pebbles.

### 2.2.2. Accretion of metal and silicate pebbles in a 2-D disk

We accrete the metal and silicate pebble groups in fixed proportions according to equations (2) and (3); that is, we assume homogeneous accretion. Furthermore, we assume constant values of disk parameters, $\Sigma_p$ and headwind velocity $v_{hw}$, and protoplanet (Earth) orbital radius and angular velocity $R$ and $\Omega$, as given in Table 3. Following Johansen and Lambrechts (2017), the accretion rate of pebbles in a 2-D disk (a thin disk with negligible thickness compared to the Hill radius) is given by

$$\frac{dM}{dt} = 2r_{acc}\Sigma_p v_a, \qquad (4)$$

where $t$ is accretion time, $M$ is the protoplanet mass, and $v_a$ is the approach velocity of pebbles. $v_a$ is the sum of the headwind velocity of a pebble $v_{hw}$, and the velocity due to Keplerian shear, $\Omega r_{acc}$ (Johansen and Lambrechts, 2017):

$$v_a = v_{hw} + \Omega r_{acc}. \qquad (5)$$

Depending on $r_{acc}$, $v_a$ and the initial mass of the protoplanet, there are two regimes of pebble accretion, often referred to as the Bondi and Hill regimes. (See Johansen and Lambrechts, 2017 for a review of both regimes). Pebble accretion transitions from Bondi to Hill regimes when the protoplanet exceeds a transition mass (Johansen and Lambrechts, 2017),

$$M_t = \frac{v_{hw}^3}{\sqrt{3}G\Omega}, \tag{6}$$

where $G$ is the gravitational constant. The Bondi regime initiates after planetesimals of ~100 km radii formed via gravitational and streaming instabilities (Johansen and Lambrechts, 2017). The present study is restricted to accretion in the Hill regime because it dominates the protoplanet's growth starting from a seed mass radius of a few hundred km or larger.

In the Hill regime, $r_{acc}$ is given approximately (Johansen and Lambrechts, 2017) by

$$r_{acc} = r_{Hill}(10St)^{1/3}. \tag{7}$$

The Hill radius $r_{Hill}$ expands with increasing protoplanet mass during accretion (Ormel and Klahr, 2010) according to

$$r_{Hill} = \left(\frac{M}{3M_{Sun}}\right)^{1/3} R, \tag{8}$$

where $M_{Sun}$ is the mass of the sun or central star. Incorporating the expression for $v_a$ from equation (5) and $r_{acc}$ from equations (7) and (8) into equation (4), we get

$$\frac{dM}{dt} = 2\Sigma_p v_{hw} R \left(\frac{10StM}{3M_{Sun}}\right)^{1/3} + 2\Sigma_p \Omega R^2 \left(\frac{10StM}{3M_{Sun}}\right)^{2/3}. \tag{9}$$

We apply equation (9) individually to the metal pebble mass $M^m$ and silicate pebble mass $M^s$ of the protoplanet as

$$\frac{dM^m}{dt} = 2\Sigma_p^m v_{hw} R \left(\frac{10St^m M}{3M_{Sun}}\right)^{1/3} + 2\Sigma_p^m \Omega R^2 \left(\frac{10St^m M}{3M_{Sun}}\right)^{2/3} \tag{10}$$

and

$$\frac{dM^s}{dt} = 2\Sigma_p^s v_{hw} R \left(\frac{10 St^s M}{3 M_{Sun}}\right)^{1/3} + 2\Sigma_p^s \Omega R^2 \left(\frac{10 St^s M}{3 M_{Sun}}\right)^{2/3}, \quad (11)$$

where

$$M = M^m + M^s. \quad (12)$$

Here, $St^m$ and $St^s$ are the Stokes numbers of the metal and silicate pebbles, respectively. The initial time, $t = 0$ stands for the initiation of the Hill regime, when the initial (i.e., seed) mass equals the transition mass as defined by equation (6). In the pebble accretion model by Johansen et al. (2021), $t=0$ corresponds to ~0.6-1.5 My for terrestrial planets, after the disk formation, consistent with the formation timeline of refractory inclusions (Krot, 2019) and early-formed chondrules (Connelly et al., 2012). Here, for $v_{hw}$ = 30 m/s, the seed mass is ~2 × $10^{-4}$ $M_E$ for our Earth model, where $M_E$ is Earth's present-day mass. The seed mass has an Earth-like (Fe+Ni)/Si=2.06. We solve equations (10)-(12) numerically using a second order Runge-Kutta method. First, we implement a grid of $St^m$ from 0.001 to 0.1 and $St^s$ from 0.001 to 0.01. We then calculate the accreted pebble masses versus time for each Stokes number combination, starting at $t = 0$, with final masses $M$ ranging from 0.01 to 10 $M_E$.

### 2.2.3. Pebble capture probability and total fluxes of drifting metal and silicate pebbles

Pebble capture probability or efficiency of pebble capture $P_p$ is the ratio of the pebble setting flux to the total flux of drifting pebbles (Ormel, 2017). The total flux of drifting pebble is the mass of pebbles crossing an orbit per unit time, given by.

$$\frac{dM_{drift}}{dt} = -2\pi R v_r \Sigma_p, \quad (13)$$

where $v_r \approx -2St v_{hw}$ is the radial velocity of a pebble (Ormel and Klahr, 2010). The negative sign signifies drift in the inward radial direction. Combining equations (13) and (9) leads to

$$P_p = \frac{1}{2\pi St^{2/3}}\left(\frac{10M}{3M_{Sun}}\right)^{1/3} + \frac{\Omega R}{2\pi v_{hw} St^{1/3}}\left(\frac{10M}{3M_{Sun}}\right)^{2/3}. \qquad (14)$$

From the above expression, we see that $P_p$ varies inversely with $St$ and is independent of $\Sigma_p$. For $P_p = 1$, settling is assured, whereas for $P_p=0$, no pebbles settle on the protoplanet. $P_p$ increases nonlinearly with $M$. We calculate pebble capture probability from equation (14) for a 0.7 $M_E$ protoplanet using the accretion parameters given in Table 3, the maximum mass of proto-Earth among our preferred models in Figure 1.

$M_{drift}$ is the sum of the masses of drifting metal pebbles $M_{drift}^m$ and silicate pebbles $M_{drift}^s$. Following equation (13), the mass flux of drifting metal pebbles is

$$\frac{dM_{drift}^m}{dt} = 4\pi R St^m v_{hw} \Sigma_p^m, \qquad (15)$$

and for silicate pebbles

$$\frac{dM_{drift}^s}{dt} = 4\pi R St^s v_{hw} \Sigma_p^s. \qquad (16)$$

As all the r.h.s. parameters are independent of time, we can solve equations (15)-(16) analytically to find the total incoming pebble flux at 1 au.

## 3. PEBBLE ACCRETION FOR EARTH

Figure 1 shows the results of our pebble accretion model for metal and silicate pebble column densities of $\Sigma_p^m = 0.095$ kg/m² and $\Sigma_p^s = 1.105$ kg/m², respectively. These partial column densities are determined using equations (2)-(3), with the nebular fractions of metal and silicate pebbles required to reproduce Earth's major element composition. Values for $X_i$ in equation (2) are taken from our best-fitting composition model CMA in Section 5.1. Numerical values of $C_i^{\text{FeNi}}$ and $C_i^{\text{Si}}$ are taken from Table 5. A total pebble column density $\Sigma_p$ = 1.2 kg/m² at 1 au is consistent with observations of evolved protoplanetary disks that are a few million years old (Bitsch et al., 2015). We chose a constant $\Sigma_p$ instead of a time-varying $\Sigma_p$ for simplicity. This is a first-order approximation to the mass balance resulting from the continuous production of chondritic component pebbles (Villeneuve et al., 2009; Connelly et al., 2012) and decreasing pebble abundance over time. The white contours in Figure 1 are total protoplanet mass, relative to Earth's mass after 2 My accretion in the Hill regime, versus $St^m$ and $St^s$, calculated using equations (10)-(12). The black contour lines in Figure 1 are the metal pebble fraction of the protoplanet after 2 My of pebble accretion. The red contour in Figure 1 corresponds to an Earth-like (Fe+Ni)/Si mass ratio of 2.06. Stokes number combinations along this curve define our preferred models with protoplanetary masses up to 2.3 $M_E$. By comparison, the full $St$ range in Figure 1 would produce protoplanet masses between 0.25 and 2.7 $M_E$.

The red dots in Figure 1 are $St$ combinations for metal grains and chondrules given in Table 1. Each dot is for a specific chondrule and a specific metal grain or metal aggregate type. The red dots on the left side are $St^m$ for individual metal grains, while those on the right side are for the maximum size possible for metal aggregates, based on the bouncing barrier. Accordingly, any $St^m$ between individual grains and the maximum for metal aggregates is possible. The turquoise box bounds all the red dots and therefore constrains the possible $St$ values (or sizes) of the chondritic component pebbles. This box also restricts the range of protoplanet masses along the red curve, (our preferred Earth models constrained by solar (Fe+Ni)/Si plus pebble size), to 0.35-0.7 $M_E$. An interesting implication of Figure 1 is that a large protoplanet (0.7 $M_E$) built with these two pebble groups would require accretion of a significant amount of metal aggregates as well as large chondrules. In contrast, a small protoplanet (0.35 $M_E$) could form by accreting mostly individual metal grains and small chondrules.

Figure 2 shows how different values of $\Sigma_p$ can lead to either a small proto-Earth or a super-Earth after 2 My of pebble accretion, using the same $x_m$ and $x_s$ as in Figure 1. Figures 2a, 2c, and 2e show protoplanet masses versus pebble accretion time for $\Sigma_p$ = 0.2, 1.2, and 2 kg/m², respectively. The curves in each plot represent a combination of $St^m$ (first number in the caption of Figure 2c) and $St^s$ (second number). The dotted lines in each figure show protoplanets of mass 0.35 $M_E$ and 0.7 $M_E$. It is evident from these plots that 0.35-0.7 $M_E$ protoplanets can form within the first 0.5 to 1.5 My of accretion in the Hill regime for $\Sigma_p$ = 2 kg/m², but a super-Earth would require a 2 My timeframe. In contrast, for $\Sigma_p$ = 0.2 kg/m², 2 My of Hill regime accretion is insufficient to build a protoplanet of 0.35-0.7 $M_E$.

Figures 2b, 2d, and 2f correspond to Figures 2a, 2c, and 2e, respectively, and show the protoplanet mass relative to Earth's mass after 2 My of Hill regime accretion, as a function of $St^m$ and $St^s$. The red dots and the turquoise boxes in each figure are the same as in Figure 1. Here, $\Sigma_p$ = 2 kg/m² leads to a super-Earth within the box, whereas $\Sigma_p$ = 0.2 kg/m² produces a very small protoplanet. In contrast, 0.35-0.7 $M_E$ protoplanets are produced for $\Sigma_p$ = 1.2 kg/m². In summary, Figure 2 indicates that a smaller or larger $\Sigma_p$ would increase or decrease the formation time, and both extremes would be unsuitable for building Earth via Hill regime accretion within 2 My. Whereas the former case would build a protoplanet on a timescale longer than the disk lifetime, the latter case would lead to super-Earth-size terrestrial planets if Hill regime growth continued for more than 1.0-1.5 My.

In Figure 3, we show pebble capture probabilities for a 0.7 $M_E$ protoplanet and the total mass of drifting pebbles relative to $M_E$ after 2 My of Hill regime accretion, as functions of $St^m$ and $St^s$, for various $\Sigma_p$, using the $x_m$ and $x_s$ as in Figure 1. This figure shows that capture probability is independent of $\Sigma_p$ and is larger for the smaller $St$ (Figures 3a, 3c, and 3e), conforming to equation (13). Capture probabilities lie between 0.22 and 0.65 inside the box. The total drifting pebble mass varies with $\Sigma_p$, as shown in Figures 3b, 3d, and 3f. It amounts to 1.8-6 $M_E$ inside the box for $\Sigma_p$ =1.2 kg/m². Therefore, for $\Sigma_p$ =1.2 kg/m², we estimate that the total pebble mass that can accrete inside Earth's orbit, the product of capture probability and total drifting pebble mass, is 1.2-1.3 $M_E$.

## 4. INVERSIONS FOR MAJOR ELEMENT COMPOSITION

### 4.1. Mass balances

Our Chondrite-based and Component-based models for deriving the Earth's major element composition are based on mass balance considerations, in which the various pebble types (chondrites or a mixture of chondrules, metal grains, AOAs, etc.) are combined in proportions that lead to best-fit matches for the bulk Earth composition. We express this mass balance by the following set of linear equations:

$$C_{ji} X_i = E_j. \qquad (17)$$

$C_{ji}$ represents the concentration of the $j^{th}$ element in the $i^{th}$ pebble type, $E_j$ is the concentration of the $j^{th}$ element in the Earth, $X_i$ is the mass fraction of the $i^{th}$ pebble type, and summation is implied over the repeated index. Given $C_{ji}$ and $E_j$ as inputs, we solve equation (17) using two different inversion techniques to find $X_i$.

### 4.1.1. Input properties

$C_{ji}$ consists of distinctly different pebble types in the two models. In the Chondrite model, elemental concentrations are for pebbles consisting of C, EL, EH, and O chondrites (Wasson and Kallemeyn, 1988) and iron meteorites (Scott, 2020). Note that each chondrite subgroup has a wide range of compositions. However, for simplicity, we are interested only in their average compositions. Here the C chondrites represent an average composition of CI, CM, CO, and CV chondrites. For O chondrites, the average is taken over L, LL, and H chondrite compositions. Groups of iron meteorites vary in their major constituents Fe (40-90 wt%) (Scott, 2020), Ni (5-18 wt% (Mittlefehldt et al., 1998), up to 60 wt% (Scott, 2020)), Co (0.3-1.3 wt%), S (0.02-8 wt%), P (0.01-2 wt%), and Co (0.3-1.3 wt%) (Scott, 2020). For simplicity, we choose a typical composition of iron meteorites from Scott (2020).

In the Component model, elemental compositions are determined using two independent datasets for the following pebble types: AOAs, CAIs, metal grains, C chondrules, E chondrules, and O chondrules. In the first data set, compositions are taken from Alexander (2022; 2019). For the metal grains, we assume 95 wt% of Fe and 5 wt% of Ni. For simplicity, we include no siderophile elements in the metal grains, which would lead to underestimates of the concentrations of minor siderophile elements. Hereafter, the model corresponding to these data is referred to as CMA. Our second data set is based on measured compositions of the individual components (Grossman et al., 1979; Ruzicka et al., 2012; Archer et al., 2014; Hezel et al., 2018b; Okabayashi et al., 2019; Kadlag et al., 2019; Van Kooten et al., 2022). Hereafter, the model corresponding to these data is referred to as CDB (Component Data Base). Like the Chondrite model, C, E, and O chondrule compositions represent mean composition over their subgroups. Additionally, we do not distinguish between FeO-poor and FeO-rich chondrules and metal grains. Compositional data for the bulk Earth are taken from Wang et al. (2018). Tables 4 and 5 contain the $C_{ji}$ for the Chondrite and the Component models, respectively. Tables S1, S2, and S3 in the Supplementary Material contain concentrations of major and minor elements in all the pebble types as well as the Earth. Note that all concentrations are expressed in mg/g.

### 4.1.2. Inversion techniques

We seek best fits to Earth's major element composition by solving equation (17) using Least-Squares (LS) and Markov-Chain Monte Carlo (MC) inversion methods. Both methods are designed to find the optimum solution for $X_i$ by minimizing a weighted root-mean-square (rms) misfit to the element concentrations. Letting $U_{E_j}$ be the observed uncertainty of the $j^{th}$ element in the Earth, we define this rms misfit as

$$\epsilon_{rms} = \sqrt{\frac{\epsilon_j \epsilon_j}{n}}, \qquad (18)$$

where $\epsilon_j = C'_{ji} X_i - E'_j$, $n$ is the total number of elements, $C'_{1i} = C_{1i}/U_{E_1}$, $E'_1 = E_1/U_{E_1}$ and so on for $j = 1, 2... n$. The main reason to use $U_{E_j}$ is to apply weight to element concentrations in proportion to their certainty. Nevertheless, LS and MC are different, because LS can converge to a rms local minimum, whereas MC should converge toward the global rms misfit (within the parameter bounds). Because of this difference, we use both methods in parallel to test for convergence and robustness. The details of the LS and MC techniques are given in the Supplementary Material.

## 5. COMPOSITION MODEL RESULTS

We fit both pebble models to Earth's most abundant major elements (Fe, Ni, Si, Mg, Ca, Al, O) and the two most abundant minor lithophile elements (Ti and Cr), which together compromise ~99% of Earth's mass.

### 5.1. Best fitting Chondrite and Component models

Figures 4 and 5 summarize the best fits for the Chondrite and Component models when we use only the seven major elements above. Here, the definition of relative misfit (RM) is $\frac{(C_{ji} X_i - E_j)}{E_j} \times 100\,\%$, which corresponds to an overestimate (underestimate) of the concentration in the model relative to Earth for a positive (negative) value of RM. The deviation of the Chondrite model from the Component model is quantified in terms of root-mean-square misfits (RMSMs = $\sqrt{\frac{RM^2}{n}}$). In Table 6, we compile the RMSMs for the fitted major elements as well as for the fitted plus the predicted element lists. The shaded blue area in each plot represents the observed uncertainty of the elemental concentrations in the bulk Earth. These values correspond to one standard deviation of the estimated concentrations as reported by Wang et al. (2018).

For the Chondrite model, the RMs of all the major elements obtained by LS (blue line) and MC (red line) lie outside the uncertainties of Earth's composition (Figure 4a). The RMs of the minor elements (Figure 4b) (which are not part of the best fit but rather calculated from the best fit for the major elements), result in anomalously high concentrations of moderately volatile elements (Na, K, S, etc.) and low concentrations for most refractory siderophile elements. However, some loss of moderately volatile elements is expected if the pebbles undergo volatilization as they settle through the hot proto-Earth atmosphere. Consequently, final abundances of the moderately volatile elements will be controlled by the protoplanet mass, thermal state of its atmosphere, and its exchange with the nebula. Similarly, for the refractory siderophile elements the misfit can be explained by an underrepresentation of these elements in our component compositions, or a higher core concentration of these elements than commonly assumed. The pie charts in Figures 4c and 4d compare the relative pebble proportions found by LS and MC, respectively. They are also tabulated in Table 6. Both techniques yield reasonably similar solutions, indicating a dominance of EL chondrites (87-90 wt%) followed by iron meteorites (6 wt%) and C chondrites (3-6 wt%).

Figures 5a and 5b are similar plots to Figures 4a and 4b but represent the results for the Component model. The dotted and solid curves in these plots correspond to the CMA and CDB models, respectively. Here, RMs for major elements are significantly improved (Figure 5a) as compared to the Chondrite model (Figure 4a). None of the RMs in Figure 5a lie outside of the uncertainty bounds for Earth except for Ni in model CDB. Although Ni is overestimated in the CDB model, it would come into agreement with the bulk Earth if we reduced our assumed Ni concentration in metal grains from 6.1 wt% to 5.5 wt%. The CMA model shows a better minor element fit as compared to the CDB model, which overestimates many of the volatile elements (Figure 5b). Even so, both models predict the minor elements better than the Chondrite model in Figure 4b. Table 6 shows that the RMSMs for both element lists are clearly larger for the Chondrite model than for the Component model. Also, it is noteworthy that both LS and MC inversions converge to nearly identical solutions for both Chondrite and Component models.

Pebble proportions in the best-fitting CDB and CMA models are shown in the pie charts in Figures 5c and 5d and in Table 6. Both models predict a large amount of C chondrules (42-74 wt%), with negligible E chondrules (0-4 wt%). Despite their elemental similarity, the CDB model predicts a large O chondrule proportion and zero CAIs, unlike the CMA model. Because CMA is our best match for major element composition among the above models, we used the chondritic component fractions in the CMA to determine the partial column densities of metal and silicate pebbles in Section 3, based on equations (2)-(3).

## 5.2. Constrained Component model

Earth is isotopically close to E chondrites for a number of elements and isotope ratios (Lodders, 2000; Javoy et al., 2010; Warren, 2011; Dauphas, 2017). In contrast, C chondrites do not match many of Earth's isotopic compositions so well. The O isotope composition of E chondrules are similar to E chondrites because the chondrules are their most abundant constituent (60-80 vol%; Brearley and Jones, 1998). Also, refractory inclusions, which are isotopically distinct from chondrules, are very rare (<0.1 vol%) in E chondrites. These observations are at odds with our best fits to Earth's bulk composition which are predominantly made from C chondrules. Therefore, it is worth examining the sensitivity of our results to various proportions of E chondrules. We do this by *a-priori* constraining E chondrules to be 60 wt% of the Component model and comparing the resulting major element concentrations to the best-fitting major element concentrations found without this constraint. We note that this high proportion of an E component is consistent with nucleosynthetic and isotope studies suggesting 50-95 wt% contribution of E chondrites to Earth's composition (Burbine and O'Brien, 2004; Dauphas, 2017; Worsham and Kleine, 2021; Savage et al., 2022; Steller et al., 2022; Martins et al., 2023). As the outputs from LS and MC inversions are virtually indistinguishable, hereafter we only show the results from LS inversions, unless stated otherwise.

Figure 6 shows results of the two Component models with the 60 wt% E chondrule constraint, with their pebble proportions given in Table 6. The RMs in Figures 6a and 6b are substantially greater for a number of elements than those of the unconstrained Component models in Figures 5a and 5b. In particular, refractory elements Ca and Al are deficient in the constrained models, because E chondrules are deficient in those elements. Also, both constrained models overestimate moderately volatile elements such as K, Zn, S, Cu, Ge, and Se by factors of 2-4, because those elements are overabundant in E chondrules compared to C chondrules (Figure 6b). The RMSMs are higher in Figure 6 than in Figure 5, which quantifies their amount of degradation (Table 6). The pie charts in Figures 6c and 6d indicate that constraining E chondrules to be 60 wt% of the total pebble mass requires large proportions of metal grains and AOAs. This indicates that these early-formed components may have been extensively incorporated in the early stages of pebble accretion, prior to the formation of chondrites. In spite of its shortcomings, the major element abundances in the CMA model with 60 wt% E chondrules are mostly within the uncertainties of estimates for the bulk Earth elemental composition.

### 5.3. Robustness of the Component model

The robustness of an inversion is a measure of stability of the inversion to changes to the input. In this section, we test the robustness of our Component model. Because this model is overdetermined, with the number of equations (elements) exceeding the total degrees of freedom (number of pebble types), no exact solution exists. Introducing more equations without changing the total number of pebble types would elevate the degree of non-uniqueness and provides a good test of model robustness. For this test, we extend our element list to include Ti and Cr, the two most abundant, non-volatile minor elements to our major element list.

Figure 7 and Table 6 summarize the results from an LS inversion with Ti and Cr included in the Component model. RMs for all the fit elements lie within the uncertainty limits, except Ni and Cr in the CDB model (Figure 7a). In addition, the predictions of minor elements shown in Figure 7b are very similar to Figure 5b. The incorporation of Ti and Cr still produces a model dominated by C chondrules for both CDB and CMA datasets. Although there are some minor differences among the other pebble types in Figures 7c and 7d compared to Figures 5c and 5d, the addition of Ti and Cr produces no major changes to the Component model.

### 5.4. Constraints from nucleosynthetic isotope data

Nucleosynthetic isotopes have proven to be a powerful constraint on the building blocks of Earth. Warren (2011) showed that the combined $\varepsilon^{54}$Cr vs. $\varepsilon^{50}$Ti ratios convincingly define an inner and outer system reservoir of material and that Earth is composed primarily of inner, non-carbonaceous feedstock. In our best fit model, we find that the best fit to the major elements is made when we have as much as 74 wt% carbonaceous chondrules, seemingly at odds with nucleosynthetic isotope data. In order to further explore this discrepancy, we determined the $\varepsilon^{54}$Cr vs. $\varepsilon^{50}$Ti composition of Earth using the best fits for the seven major elements (Fe, Ni, Si, Mg, Ca, Al, O) as the only constraint. We use $\varepsilon^{54}$Cr vs. $\varepsilon^{50}$Ti values from Warren (2011) for bulk chondrites, Torrano et al. (2024) for CAIs, and Williams et al. (2020) and Schneider et al. (2020) for AOAs and chondrules (Supplementary Material Table S4). C chondrule data (Allende and Karoonda) cluster into two populations – one with negative $\varepsilon^{54}$Cr vs. $\varepsilon^{50}$Ti values and another with positive values (Williams et al., 2020). Achondrites are among the earliest formed objects in the solar system and are expected to have a large proportion of early-formed pebbles, consistent with an early formation of proto-Earth. Given that all achondrite data have uniformly negative $\varepsilon^{54}$Cr vs. $\varepsilon^{50}$Ti values ($\varepsilon^{54}$Cr average -0.5±0.2 and $\varepsilon^{50}$Ti average -1.5±0.4; Williams et al. (2020)), we use the low-value data for C chondrules ($\varepsilon^{54}$Cr = -0.23 and $\varepsilon^{50}$Ti = -0.15) for our isotope determinations. Our mass balance results show that we have a nearly exact fit to Earth's composition when we use the components (Figures 5a-7a) rather than chondrites (Figure 4a) and notably, that our estimate for the best fit using a large proportion of C chondrules results in both $\varepsilon^{54}$Cr vs. $\varepsilon^{50}$Ti values that are indistinguishable from bulk Earth (Figure 5a). Surprisingly, our best fit for components with the high E chondrule abundance leads to a poorer fit for $\varepsilon^{54}$Cr vs. $\varepsilon^{50}$Ti (Figure 6a). Nucleosynthetic isotope data for $\varepsilon^{54}$Cr vs. $\varepsilon^{50}$Ti therefore favor the chondritic components as the building blocks of Earth with a high C chondrule proportion.

## 6. COMPARISON WITH PRIOR STUDIES

Our pebble accretion model for Earth is similar to that of Johansen et al. (2015) and Johansen et al. (2021) with several important differences. Johansen et al. (2015) consider the entire process of planetary formation, starting with planetesimal formation via streaming instability followed by embryo formation by planetesimal accretion plus chondrule accretion. Their model predicts the formation of embryos with Mars mass in ~4 My. On the other hand, Johansen et al. (2021) combine type I planetary migration and planetesimal accretion with pebble accretion of 5 terrestrial planets (including Theia) in the Hill regime. They find that proto-Earth's mass grows up to 60% in ~4 My. We focus only on proto-Earth growth via pebble accretion in the Hill regime, starting from a pre-existing seed mass. Seed mass formation, planetesimal accretion, and planetary migration are not included in our model. Also, Johansen et al. (2021) consider an exponentially decaying pebble column density during the accretion period. In contrast, we assume a constant pebble column density due to the cumulative effects of pebble production and reduction in the disk, which leads to a faster growth rate in our model, compared to Johansen et al. (2021). Continual formation of chondrules is supported by age determinations for individual chondrules with span several million years (e.g., Villeneuve et al., 2009). It is noteworthy that the present study also includes the accretion of refractory inclusions and metal grains in addition to chondrules. Nevertheless, our prediction of 0.3-0.7 $M_E$ proto-Earth formation in 2 My in the Hill regime, while slightly faster, is still in broad agreement with the above models.

In terms of formation via pebble accretion, there are two important differences between previous pebble models and ours. First, unlike prior work, we accrete the nebular fractions of two pebble groups: metal and silicate in proportion to the solar (Fe+Ni)/Si ratio of 1.84 (Lodders, 2000). In addition, our best-fitting models match Earth's (Fe+Ni)/Si ratio. Therefore, our pebble model reproduces this property of Earth's composition in a way that is consistent with the nebular composition. Second, we use Stokes numbers of all chondritic components based on their measured sizes and densities (Table 1). Previous studies focusing on chondrule accretion assume a range of chondrule sizes, and they do not build planets by using just the measured $St$ values of chondritic components. In contrast, our best-fitting models successfully reproduce Earth's bulk composition via accretion of pebbles consisting of observed chondritic components, within the solar disk lifetime.

Multiple previous studies have proposed matches to Earth's composition using mixtures of chondrites (Javoy, 1995; Burbine and O'Brien, 2004; Javoy et al., 2010) or chondritic components (Anders, 1977; Alexander, 2022) as sources. One of the major findings of the present study is that component sources dominated by C chondrules better account for Earth's major element composition compared to sources dominated by E chondrites. Alexander (2022) also found that C chondrules were dominant in his bulk Earth models. In contrast, Javoy (1995) argued that Earth's isotope composition can be reproduced using EH chondrites alone, while Javoy et al. (2010) match Earth's bulk composition by allowing the bulk composition of E chondrites to vary within a large range of uncertainty.

Furthermore, we find no solution to any combination of chondritic bodies that matches Earth's major element composition. Burbine and O'Brien (2004) attempted to match Earth's O isotopic and bulk chemical compositions from mixtures of 13 C and NC chondrites. Their best-fitting mixture on average contains 55 wt% of EL and 26 wt% of O (H+L+LL) chondrites, which differs from the ~90 wt% of EL chondrites in our Chondrite model. Table 7 and Figure 8 compare the model composition of Burbine and O'Brien (2004) and our Chondrite and Component (CDB and CMA) models. Despite the differences in chondritic mixtures, the relative misfits of Fe, Si, Mg, and Al in the Burbine and O'Brien (2004) model show a similar trend to our Chondrite model, particularly Si (Figure 8). However, both of their models fail to match Earth's major elemental abundances within observed uncertainties. More significantly, our Component models outperform all the chondrite-based models in Figure 8.

An early attempt to make Earth's composition out of chondritic components (refractory condensates, metal, and silicate) was made by Anders (1977), who assumed that the Earth was formed in parallel to chondritic bodies from the same source materials. A recent study by Alexander (2022) takes a similar approach, that is, reproducing the bulk Earth composition out of metal, silicate chondrules, and refractory inclusions (CAI, AOA) with the addition of a late veneer component. Comparisons between the relative pebble proportions and model compositions in our CDB and CMA models with the above studies are shown in Table 7. To make comparisons between these models, we have combined several of the individual components from each model into three major components: refractory inclusions, metal, and chondrules. Chondrules are the major component followed by large metal proportions in CDB and Anders, (1977), compared to the other models. The reason for the large metal proportion in Anders (1977) is his exclusion of metal parts from silicate chondrules. Our CMA model pebble mixture is similar to Alexander (2022) because it is based on the same data, except for metal grains. Accordingly, Table 7 shows an excess of Ni in the latter model due to a higher Ni content in C-like metal grains (by ~1.3 wt%). This is also true of the CDB model. In Figure 8, we include the relative misfits of the major elements for the Anders (1977) and Alexander (2022) models. Two takeaways from Figure 8 are: (i) our CMA model is the best among all six in terms of overall fit, and (ii) the Anders (1977), Burbine and O'Brien (2004) and our Chondrite model are substantially outside the uncertainty range for the bulk Earth.

Our CDB model is novel in several respects. The input composition represents the mean composition of metal grains, refractory inclusions, and chondrules. In contrast, our CMA model is similar to Alexander (2022) in that it relies on input compositions that were devised to satisfy *a-priori* requirements. In particular, some of Alexander's compositions and isotope ratios are derived from extrapolations of elemental ratios or inferences (e.g., O isotope composition of chondrules), rather than actual measured data. This approach results in an excellent fit, but partly because of the *a-priori* assumptions that were made. In contrast, our CDB model solely relies on measured data from chondrites.

## 7. DISCUSSION

**7.1. Implications of our pebble accretion model for terrestrial planets:** We showed in Section 3 that 0.35-0.7 $M_E$ protoplanets can form in approximately 2 My or less by Hill regime accretion of pebbles with the observed sizes of chondritic components: metal grains, chondrules, and refractory inclusions and disk parameters in Table 3. This formation time refers to a time after the onset of Hill regime accretion, not time based on CAI formation. However, considering that the onset of the Hill regime occurs at ~0.6-1.5 My after disk formation for proto-Earth and Theia, as proposed by Johansen et al. (2021), the total accretion time for the proto-Earth and Theia is 2.6-3.5 My. This timeline implies that the Earth formed early relative to most chondritic bodies and well within the expected lifetime of the solar nebula.

In addition to Earth, our pebble accretion model has relevance to other terrestrial planets. In Section 3, we estimated that for a pebble column density of 1.2 kg/m$^2$, the total pebble mass that can accrete inside Earth's orbit, the product of capture probability and total drifting pebble mass, is 1.2-1.3 $M_E$ in 2 My of Hill regime accretion. However, for a higher pebble column density of 1.8 kg/m$^2$, 2.5-9 $M_E$ pebble mass can reach Earth's orbit in 2 My of Hill regime accretion. Accordingly, up to ~2 $M_E$ pebbles can accrete inside Earth's orbit from chondrule, refractory inclusions, and metal grains, for a pebble column density of 1.8 kg/m$^2$. Under such a solar disk condition, ~2 My of accretion in the Hill regime is enough to build Earth (proto-Earth and Theia), Moon, Mercury, and Venus allowing all to form within the protoplanetary disk lifetime.

An important consideration is the blocking capacity of Jupiter to inward pebble migration. The rapid growth of Jupiter would have created a barrier to inward pebble drift, reducing the time-integrated pebble mass for the terrestrial planets derived from material originating beyond Jupiter's orbit (Morbidelli and Raymond, 2016). For this reason, it is important to determine what part of our estimated total pebble mass of 1.8-6 $M_E$ arrives at Earth's orbit from beyond Jupiter's orbit. Our Earth model with a constant pebble column density ($\Sigma_p$ = 1.2 kg/m$^2$) predicts that the origin of 1.8-6 $M_E$ pebbles is ~11-21 au. Using instead a radially varying $\Sigma_p$ in a standard disk model for super-Earths (Johansen and Lambrechts, 2017), the origin is ~11-33 au. Both models indicate that most Earth-building pebbles must cross Jupiter's orbit in order to reach Earth's orbit. However, there are caveats here. First, we implemented the simplest pebble accretion model for Earth without considering the formation timeline of chondritic components. The radiometric studies show that the age distribution of chondrules is between 1 and ~4 My after CAIs (Villeneuve et al., 2009; Pape et al., 2019) or even earlier, starting at ~0 My (Connelly et al., 2012). Therefore, if we consider that pebbles consisting of chondritic components were continuously forming in the terrestrial feeding zone simultaneously with pebble accretion, we still can build 0.35-0.7 $M_E$ protoplanet from the pebbles inside Jupiter. Equally importantly, a recent study shows that Jupiter does not capture all dust aggregates, allowing some aggregates to enter the terrestrial feeding zone from beyond Jupiter (Johansen et al., 2021). These dust aggregates may grow into larger pebbles (as evidenced by the younger chondrule ages) which can later accrete into protoplanets. Recently, Brasser (2024) have shown that 10% to 25% of the mass from the Jupiter-Saturn accretion region can enter the inner solar system in the first 5 My. Moreover, Lambrechts et al. (2019) showed that, even at a low pebble flux due to the presence of Jupiter, ~100 $M_E$ pebble mass can enter the inner disk, which can reproduce the terrestrial masses. While only ~2 $M_E$ would accrete to build the terrestrial bodies, the rest of the pebbles would fall onto the Sun. In fact, it can be argued that if Jupiter had not been there to block the unimpeded addition of pebbles to the inner solar system, then the inner solar system bodies would have been far larger than they are.

**7.2.   Chondrites versus chondritic components as Earth's building blocks:**

The classical chondrite model attempts to mix chondrites in varying proportions to make an Earth-like composition. Iron meteorites are generally not included in such models because they are fragmented cores of planetesimals/protoplanets that formed before chondrites (Kelly et al., 1977; Goldstein et al., 2009). However, in Section 5, our Chondrite model includes iron meteorites as a source for the following reasons. Planetesimals can fragment via mutual collisions and generate dust and pebbles (Gerbig et al., 2019). Iron meteorites themselves are the consequences of such collisions. Under this assumption, fragments of iron meteorite and chondrites qualify as pebbles. When we attempted to match Earth's major element composition with a mixture of only chondrites, we found that such a model is depleted in metals and refractory siderophiles more than our Chondrite model that includes iron meteorite (Figure 4). Adding iron meteorites as a component in our chondrite model improves the fit, but even with the addition of iron meteorites, the overall fit is poor.

Section 5 shows that mixtures of chondritic components far better match Earth's major element composition than chondrites. There is a fundamental difference in the Earth formation process when components rather than chondrites are used. Section 2.1.1 shows that pebble accretion naturally filters out the matrix among all the chondritic components. However, laboratory experiments (Ormel et al., 2008; Beitz et al., 2012) have shown how dust particles can stick on the surface of pebbles, which explains the mechanism of matrix-rim formation around the chondrules. Therefore, the matrix can co-accrete onto a growing proto-Earth along with chondrules to some extent (Ormel et al., 2008). Matrix dust can also stick together and potentially reach high enough Stokes numbers to be accreted efficiently. In this regard, we included a mean matrix composition from (Hezel et al., 2018b) in the CDB model as an additional pebble type. The best fits of this model to Earth's major element composition find 23.88 wt% metal grains, 44.44 wt% C chondrules, 0 wt% E chondrules, 23.94 wt% O chondrules, 0 wt% CAI, 6.13 wt% AOA, and 1.61 wt% matrix. This solution shows no difference from the original CDB model that excludes a matrix (Figure 5, Table 6). Collectively, chondritic components far better match Earth's composition than chondrites. This is because our Component model allows the component proportions to vary while their fractions are predetermined in the chondrites.

**7.3. Secular variation of the chondritic components:** CAIs (Connelly et al., 2012; Krot, 2019) and AOAs (Krot et al., 2009) are the oldest solids condensed from the solar nebula. However, the timing of chondrule formation remains unsolved. Some studies suggest that chondrule formation started as late as 1.8 My after CAI formation (Pape et al., 2019) resulting in a chondrule time gap (Villeneuve et al., 2009; Pape et al., 2019; Krot, 2019; Piralla et al., 2023). Others have no such time gap, with chondrule formation starting contemporaneously with the CAIs and lasting over ~3-4 My (Connelly et al., 2012; Bizzarro et al., 2017; Bollard et al., 2017;). Metal grains most likely condensed during or before the silicate chondrules in a dense nebula (Grossman, 1972; Campbell et al., 2002). The lack of early-formed chondrules in chondrites may simply be a manifestation of the fact that chondrites formed at a later time. Evidence for older chondrule incorporation in terrestrial planetesimals may be lacking because the early-formed bodies that would have incorporated them were heated ($^{26}$Al decay) to the point of melting, thereby eliminating evidence of their existence.

The formation of refractory inclusions was restricted to the inner solar system. They formed at <0.1 au (Bizzarro et al., 2017; Krot, 2019) in less than ~0.2 My. Their abundance is much less than 1 vol% for all NC chondrites but as much as 10-13 vol% for C chondrites (Weisberg et al., 2006). Refractory inclusions were presumably brought to the outer solar system by turbulent diffusion and preserved by the pressure gap created by Jupiter (Desch et al., 2018). Inward of this gap, refractory inclusions would have been drawn into the Sun by aerodynamic drag or incorporated into the early formed planetesimals or embryos that ultimately became the terrestrial planets. However, the embryos for the terrestrial planets, if formed early, would have incorporated the inner solar system refractory inclusions by the process of pebble accretion.

Figure 9 depicts a simple schematic of the secular evolution of terrestrial planet formation. The pristine chemical composition and age distributions of CAIs, AOAs, chondrules, and metal grains require them to be free-floating objects before their accretion. In reality, the secular variation of the chondritic component age would lead to a heterogeneous accretion for Earth (Grossman, 1972). This is evident from our Component models where Earth's major element composition is satisfied by a higher proportion of refractory inclusions (AOA+CAI) than their nominal proportions in NC chondrites. For instance, our best inversion model, CMA, requires ~15 wt% of refractory inclusions (Table 6). By comparison, they are found in their largest abundance of 13 vol% in CO chondrites (Scott and Krot, 2003). Additionally, the AOA proportion in our Component models is 6-23 wt% which seems unexpectedly high but consistent with the observation that the AOAs are the most abundant refractory inclusions in most C chondrites (Torrano et al., 2024).

The first phases of pebble accretion would naturally incorporate the early-formed material in the nebula, specifically, refractory inclusions (CAIs + AOAs) and metal grains. This is shown in Figure 9a. The high proportion of refractory inclusions in the early inner nebula (Figure 9a) were not assimilated in the later-formed NC chondrites (Figures 9b and 9c). Evidence for high proportions of early-formed refractory materials and chondrules is lacking in NC chondrites because they formed after most refractory grains were either drawn into the Sun or incorporated into the nascent terrestrial planets. This is shown in Figures 9b and 9c when the chondrules continue to form and accrete into the proto-Earth. The selective incorporation of early materials into the growing Earth (as opposed to chondrite parent bodies) might also explain the enrichment of *s*-process isotopes in Earth compared to chondrites (Mezger et al., 2020).

**7.4. Volatile elements on Earth:** Our major element fits do not consider highly volatile elements such as C and S because our inversion methods do not allow for element-specific mass loss. Nevertheless, because volatile element loss due to outgassing and evaporation are expected from pebble accretion (Johansen et al., 2023b), it is of interest to compare the abundances and trends of volatile elements predicted by our various models.

On average, our CDB models overestimate the moderately volatile elements (Figures 5b, 6b, and 7b). In contrast, our unconstrained CMA model in Figure 5b slightly underestimates the moderately volatile elements. This difference is mostly due to the compositional data used in each model type. Our CDB models use the measured abundances of volatiles in chondritic components, whereas our CMA models use volatile abundances from Alexander (2022) that follow the same trends as Earth's present-day inventory. Accordingly, the overabundance of moderately volatile elements in our CDB models compared to our CMA models is consistent an over-estimate of volatiles, which is explained by volatile loss during accretion.

**7.5. O isotope composition of Earth:** Although this paper focuses on Earth's elemental composition, it is important to place our models within the constraints of O isotopes. It has been shown that no mixture of solar system materials matches the Earth's chemical and isotopic composition simultaneously (Alexander, 2022). Although we were successful in reproducing Earth's O isotope composition with a mixture of CV, CM, and L chondrites, our inversion yields a best fit of 3 wt% CV, 27 wt% CM, and 70 wt% L, which fails to match Earth's major element composition within the uncertainties. Even with the addition of EH or EL chondrites, the major element misfit with O isotope is worse than our Chondrite model with no isotopic constraints (Figure 4a). On the other hand, our Component models suggest that a large proportion of C chondrules and refractory inclusions but negligible E chondrules would be incompatible with Earth's O isotope composition. In general, C chondrules are somewhat isotopically lighter for $\Delta^{17}O$ values than Earth, although some C chondrules do have positive $\Delta^{17}O$ values (Williams et al., 2020). Our attempt to match Earth's O isotope composition with only E, O, and C chondrules yields 90.65 wt% E chondrules + 9.35 wt% C chondrules and produces a far worse major element fit than all of our Component models. Our CDB model in Section 5.1 predicts $\delta^{18}O$ and $\delta^{17}O$ of -0.16 ‰ and –2.9 ‰ for the Earth, respectively. For comparison, our CMA model predicts -6.6 ‰ and -9.7 ‰, respectively. Alexander (2022) suggested several explanations for the misfit, including that the $\Delta^{17}O$ value ($\Delta^{17}O = \delta^{17}O - 0.528\times\delta^{18}O$) of chondrules was as high as 5‰. Alternatively, he suggests that refractory inclusions may have been reprocessed during chondrule formation by exchange with a heavy nebular gas. However, we suspect that the matrix, owing to heavier $\delta^{18}O$ and $\delta^{17}O$ than Earth and other chondritic components e.g., refractory inclusions, could bridge the gap between major element and O isotope compositions in the component models.

Our CDB model in Section 7.2 that includes the matrix fits Earth's O isotopes only if the matrix has $\delta^{18}O$ and $\delta^{17}O$ values as high as 346 ‰ and 293 ‰, respectively. Such highly anomalous values are indeed found in some primitive grains embedded in the matrix (Barosch et al., 2022). In light of this, we solve for a best-fit to Earth's O isotope composition by fixing the matrix proportion at 3 wt%. This procedure yields 36 wt% C chondrules, 58 wt% E chondrules, 1 wt% O chondrules, 2 wt% CAI, and 3 wt% AOA, with predicted $\delta^{18}O$ and $\delta^{17}O$ matrix values of 51.4 ‰ and 17.62 ‰, respectively. Unfortunately, this model fails to match Earth's major element composition. In summary, we are unable to find a single model that satisfies Earth's major element and O isotope composition simultaneously. Table 8 summarizes the compatibility/incompatibility of various models with Earth's major element and O isotope composition.

Our model results are based on the compositions of sampled meteorites. However, some interpretations of nucleosynthetic anomalies require that, during its formation, Earth incorporated "missing components", i.e., materials with isotopic properties that are distinct from sampled meteorites (Fischer-Gödde and Kleine, 2017; Mezger et al., 2020; Burkhardt et al., 2021). It has been proposed that these missing components probably originated in the innermost part of the solar system (Mezger et al., 2020). We predict that the missing component(s) possibly can have the matrix-like major element composition or an early, as yet unsampled refractory phase and an anomalously high O isotope.

## 8. SUMMARY

1) Mixtures of chondritic components consisting of metal grains, chondrules, and refractory inclusions match Earth's major element composition within uncertainties; mixtures of various chondrites and iron meteorites do not.

2) We find that C chondrules contribute substantially to Earth's major element and nucleosynthetic $\varepsilon^{54}Cr$ and $\varepsilon^{50}Ti$ compositions. Additionally, our Chondrule models indicate a greater proportion of early condensates (CAIs + AOAs) than found in NC chondrites. This is explained by their accretion into the proto-Earth before the start of chondrite formation.

3) Approximately two million years of pebble accretion in the Hill regime builds 0.35-0.7 $M_E$ protoplanets from pebbles having the observed sizes of chondritic components: metal grains, chondrules, and refractory inclusions. An even shorter time span is possible if the pebble column density was higher than 1.2 kg/ m² in the early solar system.

4) Constraints on the protoplanet metal pebble fraction from the solar (Fe+Ni)/Si ratio implies that a large proto-Earth (>0.7 $M_E$) requires more metal grain aggregates and larger chondrules than that of a smaller proto-Earth (<0.35 $M_E$).

**CRediT authorship contribution statement**

**Susmita Garai:** Conceptualization, Methodology, Software, Validation, Formal analysis, Investigation, Visualization, Data curation, Writing – original draft, Writing – review & editing. **Peter L. Olson:** Conceptualization, Methodology, Validation, Supervision, Writing – review & editing. **Zachary D. Sharp:** Conceptualization, Methodology, Funding acquisition, resources, Validation, Supervision, Writing – review & editing.

**Acknowledgment**

This work was supported by EAR1953992 grant by National Science Foundation. The authors would like to thank Eric Lindsey for his assistance with inversion methods. The thoughtful comments by reviewers, especially the incorporation of nucleosynthetic $^{54}$Cr and $^{50}$Ti helped improve the manuscript significantly. Additionally, Conel Alexander provided numerous detailed comments.  We are thankful to the reviewers.

**Appendix A: Supplementary Material**

The Supplementary Material related to this article includes 1) a figure showing pebble streamlines for varying pebble size, 2) details of the implementation of Least-Squares and Markov-chain-Monte-Carlo inversion techniques, 3) the elemental compositions, and 4) the O, Ti, and Cr isotopic compositions of chondritic components, chondrites, and bulk Earth.

Table 1: Stokes numbers of various chondritic components based on their observed densities and radii.

| Chondritic component | Parent body | Mean density $\rho_p$ (g/cm$^3$) | Mean radius $r_p$ (mm) | Estimated $St$ |
|---|---|---|---|---|
| Metal grains | Kelly (K) | 6.098[b] | 0.124[b] | 0.002 |
|  | Bjurbole (B) | 6.362[b] | 0.134[b] | 0.002 |
|  | Hammond Down (H) | 6.750[b] | 0.138[b] | 0.002 |
| Metal aggregates | Kelly (K) | 6.098[b] | 3.396 | 0.052 |
|  | Bjurbole (B) | 6.362[b] | 3.348 | 0.053 |
|  | Hammond Down (H) | 6.750[b] | 3.282 | 0.056 |
| C chondrules | CV | 2.32[d] | 0.450[a] | 0.003 |
|  | CV3 | 2.467[d] | 0.460[d] | 0.003 |
| E chondrules | EL | 3.2[#] | 0.250[a] | 0.002 |
|  | EH | 3.2[#] | 0.115[a] | 0.001 |
| O chondrules | H | 3.195[c] | 0.225[a] | 0.002 |
|  | L | 3.262[c] | 0.250[a] | 0.002 |
|  | LL | 3.05[c] | 0.275[a] | 0.002 |
|  | H3 | 3.05[f] | 0.490[f] | 0.004 |
|  | H5 | 2.585[d] | 0.36[d] | 0.002 |
|  | L3 | 3.12[f] | 0.412[f] | 0.003 |
|  | L4 | 2.97[d] | 0.364[d] | 0.003 |
|  | LL3 | 3.027[f] | 0.603[f] | 0.005 |
| CAI | CO, CM, CV, CR | 3.5[g,h] | 0.1-10[e] | 0.0009-0.088 |
| AOA | CO, CV, CR | 3.3[*] | 0.1-5[e,h] | 0.0008-0.041 |
| Matrix | CI, E, O | 2.7[c] | <0.005[e,i] | <10$^{-5}$ |

Sources: a Friedrich et al. (2015); b Kuebler et al. (1999); c Gooding (1983); d Teitler et al. (2010); e Scott and Krot (2003); f Gooding (1983); g Clayton and Mayeda (1977); h Charnoz et al. (2015); i Hezel et al. (2018a) ; * Assuming a Forsterite-like composition; # Assuming an enstatite-like composition.

Table 2: Pebble accretion variables.

| Property | Notation | Unit |
| --- | --- | --- |
| Protoplanet mass | $M$ | kg |
| Metal pebble mass of the protoplanet | $M^m$ | kg |
| Silicate pebble mass of the protoplanet | $M^s$ | kg |
| Total flux of drifting pebbles at 1 au | $M_{drift}$ | kg |
| Total flux of drifting metal pebbles 1 au | $M^m_{drift}$ | kg |
| Total flux of drifting silicate pebbles 1 au | $M^s_{drift}$ | kg |
| Accretion radius | $r_{acc}$ | au |
| Hill radius | $r_{Hill}$ | au |
| Accretion time, Hill regime | $t$ | year |
| Pebble capture probability | $P_p$ | -- |
| Approach velocity of a pebble | $v_a$ | m/s |
| Elemental concentrations | $C$ | mg/g |
| Pebble fraction | $X$ | -- |
| Metal pebble fraction | $x_m$ | -- |
| Silicate pebble fraction | $x_s$ | -- |

Subscripts:

$drift, acc, Hill, p, a$ = drifting, accretion, Hill, pebble, approach.

Superscripts and Superscripts:

$m, s$ = metal, silicate.

Table 3: Disk and protoplanet parameters and their numerical values used in this study.

| **Property** | **Notation** | **Value(s)** |
|---|---|---|
| Earth's mass | $M_E$ | 5.972 x $10^{24}$ kg |
| Solar mass | $M_{Sun}$ | 1.989 x $10^{30}$ kg |
| Stokes number of metal pebbles | $St^m$ | $10^{-3}$-$10^{-1}$ |
| Stokes number of silicate pebbles | $St^s$ | $10^{-3}$-$10^{-2}$ |
| Transition mass | $M_t$ | ~$10^{-3}\,M_E$ |
| Gravitational constant | $G$ | 6.67 x $10^{-11}$ m³/kg s² |
| Orbital radius of Earth | $R$ | 1.496 x $10^{11}$ m |
| Orbital speed of Earth | $\Omega$ | 1.99 x $10^{-7}$ radian/s |
| Headwind velocity | $v_{hw}$ | 30 m/s |
| Column density of gas | $\Sigma_g$ | 1000 kg/m² |
| Total column density of pebbles | $\Sigma_p$ | 0.2, 1.2, 2 kg/m² |
| Partial column density of metal pebbles | $\Sigma_p^m$ | 0.016, 0.095, 0.158 kg/m² |
| Partial column density of silicate pebbles | $\Sigma_p^s$ | 0.184, 1.105, 1.842 kg/m² |

Subscripts:

$E, Sun, t, hw, g$ = Earth, Sun, transition, headwind, gas.

Superscripts:

$m, s$ = metal, silicate

Table 4: Input concentrations used for the Chondrite model, in mg/g.

| Element | IM[a] | C Chon[b] | EL Chon[b] | EH Chon[b] | O Chon[b] | Earth[c] | Earth uncertainty[c] |
|---|---|---|---|---|---|---|---|
| Fe | 910 | 218 | 290 | 220 | 225 | 312 | 10 |
| Ni | 80 | 12.5 | 17.5 | 13 | 12.73 | 17.7 | 0.8 |
| Si | 2 | 137 | 167 | 186 | 181 | 160 | 8 |
| Mg | 0 | 126 | 106 | 141 | 147 | 151 | 2 |
| Ca | 0 | 14.1 | 8.5 | 10.1 | 12.9 | 16.7 | 1 |
| Al | 0 | 13.0 | 8.1 | 10.5 | 11.8 | 15.4 | 0.9 |
| O | 0 | 408 | 280 | 310 | 378 | 308 | 7 |
| Ti | 0.005 | 0.69 | 0.45 | 0.58 | 0.61 | 0.792 | 0.05 |
| Cr | 0.02 | 3.21 | 3.15 | 3.05 | 3.76 | 3.88 | 0.58 |

IM= Iron Meteorite; Chon = Chondrite.

Sources: a Scott (2020); b Wasson and Kallemeyn (1988); c Wang et al. (2018).

Table 5: Input concentrations used for the Component models, in mg/g.

| Dataset | Element | MG[a,b,c] | C Ch[d] | E Ch[d] | O Ch[d] | CAI[e] | AOA[f,g] |
|---|---|---|---|---|---|---|---|
| CDB | Fe | 908 | 115 | 27.3 | 99.9 | 40.3 | 60.8 |
| | Ni | 61.5 | 6.95 | 1.02 | 3.46 | 0.19 | 2.6 |
| | Si | 6.70 | 206 | 259 | 232 | 102 | 180 |
| | Mg | 0.09 | 210 | 206 | 171 | 40.8 | 226 |
| | Ca | 0.28 | 20.8 | 10.1 | 18 | 95.4 | 32.9 |
| | Al | 0 | 20.5 | 11.4 | 17.3 | 202 | 38.8 |
| | O | 0 | 385 | 459 | 437 | 388 | 415 |
| | Ti | 0 | 1.37 | 0.48 | 1.32 | 2.53 | 0.41 |
| | Cr | 1.41 | 3.85 | 2.49 | 3.81 | 0.60 | 1.66 |

| Dataset | Element | MG | C Ch[h] | E Ch[i] | O Ch[i] | CAI[h] | AOA[h] |
|---|---|---|---|---|---|---|---|
| CMA | Fe | 950 | 269 | 270 | 227 | 0 | 47.8 |
| | Ni | 50 | 15.5 | 16.1 | 12.9 | 0 | 2.23 |
| | Si | 0 | 180 | 185 | 193 | 118 | 190 |
| | Mg | 0 | 159 | 124 | 153 | 60.4 | 230 |
| | Ca | 0 | 14.8 | 9.88 | 13.4 | 193 | 10.5 |
| | Al | 0 | 14 | 9.3 | 12.4 | 183 | 16 |
| | O | 0 | 332 | 320 | 352 | 432 | 409 |
| | Ti | 0 | 0.75 | 0.51 | 0.66 | 9.85 | 0.76 |
| | Cr | 0 | 4.07 | 3.28 | 4.02 | 0.41 | 1.18 |

MG = Metal grain; Ch = Chondrule.

Sources: a Van Kooten et al. (2022); b Kadlag et al. (2019); c Okabayashi et al. (2019); d Hezel et al. (2018b); e Archer et al. (2014); f Ruzicka et al. (2012); g Grossman et al. (1979); h Alexander (2022); i Alexander (2019).

Table 6: Summary of the inversion results. Pebble proportions are given in wt%.

| | | | Chondrite model | | | | | |
|---|---|---|---|---|---|---|---|---|
| | **Inversion** | **IM** | **C Chon** | **EL Chon** | **EH Chon** | **O Chon** | **RMSM$^\Lambda$** | **RMSM$^{\Lambda\Lambda}$** |
| | **LS** | 6.67 | 5.59 | 87.38 | 0 | 0 | 0.1949 | 1.27 |
| | **MC** | 6.07 | 3.48 | 90.16 | 0.05 | 0.03 | 0.1964 | 1.27 |

| | | | Component models | | | | | |
|---|---|---|---|---|---|---|---|---|
| **Model** | **Inversion** | **MG** | **C Ch** | **E Ch** | **O Ch** | **CAI** | **AOA** | **RMSM$^\Lambda$** | **RMSM$^{\Lambda\Lambda}$** |
| **CDB** | **LS** | 24.23 | 42.09 | 0 | 27.10 | 0 | 6.58 | 0.041 | 0.62 |
| | **MC** | 24.27 | 41.4 | 0.08 | 28.52 | 0.03 | 5.67 | 0.041 | 0.63 |
| **CMA** | **LS** | 11.26 | 73.64 | 1.61 | 0 | 2.85 | 10.83 | 0.0087 | 0.24 |
| | **MC** | 11.89 | 67.35 | 3.91 | 1.27 | 3.10 | 12.48 | 0.0084 | 0.22 |

| | | | Constrained Component models | | | | | |
|---|---|---|---|---|---|---|---|---|
| **Model** | **Inversion** | **MG** | **C Ch** | **E Ch** | **O Ch** | **CAI** | **AOA** | **RMSM$^\Lambda$** | **RMSM$^{\Lambda\Lambda}$** |
| **CDB** | **LS** | 28.13 | 0 | 60 | 0 | 2.30 | 9.57 | 0.12 | 0.7 |
| | **MC** | 28.43 | 0 | 60 | 0 | 2.35 | 9.21 | 0.12 | 0.7 |
| **CMA** | **LS** | 12.94 | 0 | 60 | 0 | 4.77 | 22.29 | 0.06 | 1.1 |
| | **MC** | 11.43 | 0.01 | 60 | 0 | 5.03 | 23.52 | 0.06 | 1.1 |

| | | | Robustness of Component models | | | | | |
|---|---|---|---|---|---|---|---|---|
| **Model** | **Inversion** | **MG** | **C Ch** | **E Ch** | **O Ch** | **CAI** | **AOA** | **RMSM$^\Lambda$** | **RMSM$^{\Lambda\Lambda}$** |
| **CDB** | **LS** | 25.39 | 31.67 | 8.05 | 22.34 | 0 | 12.56 | 0.098 | 0.57 |
| | **MC** | 25.50 | 29.66 | 7.97 | 24.45 | 0 | 12.42 | 0.098 | 0.59 |
| **CMA** | **LS** | 9.91 | 79.97 | 0 | 0 | 2.28 | 7.84 | 0.043 | 0.25 |
| | **MC** | 10.13 | 78.63 | 0.1 | 0.23 | 2.4 | 8.51 | 0.044 | 0.25 |

IM= Iron Meteorite; Chon = Chondrite; MG = Metal grain; Ch = Chondrule; LS = Least-Squares inversion; MC = Markov-Chain Monte Carlo inversion; RMSM = root-mean-square misfit. RMSM$^\Lambda$ is calculated for only fit elements. RMSM$^{\Lambda\Lambda}$ is calculated for all 35 elements.

Table 7: Comparison of pebble proportions (wt%) in four component models and comparison of major element (mg/g) plus $\varepsilon^{54}Cr$ and $\varepsilon^{50}Ti$ compositions in four component models, two chondrite models, and Earth.

| Pebble proportions (wt%) of four component models | | | | |
|---|---|---|---|---|
| **Pebble type** | **CDB** | **CMA** | **Anders (1977)** | **Alexander (2022)** |
| Metal | 24.23 | 11.26 | 36 | 11.74 |
| Chondrules | 69.19 | 75.25 | 53.2 | 73.53 |
| Refractory Inclusions* | 6.58 | 13.68 | 9.2 | 13.15 |

| Major element composition (mg/g) plus $\varepsilon^{54}Cr$ and $\varepsilon^{50}Ti$ of four component and two chondrite models | | | | | | | |
|---|---|---|---|---|---|---|---|
| **Element** | **CDB** | **CMA** | **Anders (1977)** | **Alexander (2022)** | **Chondrite model** | **Burbine and O'Brien (2004)** | **Earth (uncertainties) (Wang et al., 2018)** |
| Fe | 302.5 | 313.6 | 358.7 | 313.7 | 281.2 | 295.1 | 312 (10) |
| Ni | 19.12 | 17.53 | 20.4 | 19.38 | 19.27 | - | 17.7 (0.8) |
| Si | 164.7 | 160.3 | 143.4 | 159.7 | 188.5 | 187.6 | 160 (8) |
| Mg | 151.6 | 150.7 | 132.1 | 150.6 | 144.3 | 140.1 | 151 (2) |
| Ca | 16.03 | 16.42 | 19.3 | 16.33 | 10.68 | - | 16.7 (1) |
| Al | 16.02 | 15.54 | 17.7 | 15.46 | 10.99 | 11.8 | 15.4 (0.9) |
| O | 310.8 | 306.8 | 285 | 308.1 | 325.9 | - | 308 (7) |
| RMSM | 0.041 | 0.009 | 0.133 | 0.038 | 0.195 | 0.152 | 0.006 |
| $\varepsilon^{54}Cr$ | 0.017 | -0.063 | - | - | 0.114 | - | 0.08** |
| $\varepsilon^{50}Ti$ | -0.042 | 0.042 | - | - | 0.319 | - | 0** |

* Refractory Inclusions =AOAs + CAIs. ** Warren (2011). RMSM = root-mean-square misfit calculated for major elements.

Table 8: Compatibility/Incompatibility of various models with Earth's major element and O isotope composition.

| Sources | Earth's major element composition | Earth's O isotope composition |
|---|:---:|:---:|
| 1. Chondrules, metal grains, and refractory inclusions | ✓ | ✗ |
| 2. Chondrites | ✗ | ✓ |
| 3. Chondrules, metal grains, and refractory inclusions + matrix material | ✓ | ✗ |
| 4. Chondrules, metal grains, and refractory inclusions + matrix material with anomalously high O isotope composition | ✓ | ✓ |
| 5. Chondrules, metal grains, and refractory inclusions + missing source (?) | ✓ | ✓ |

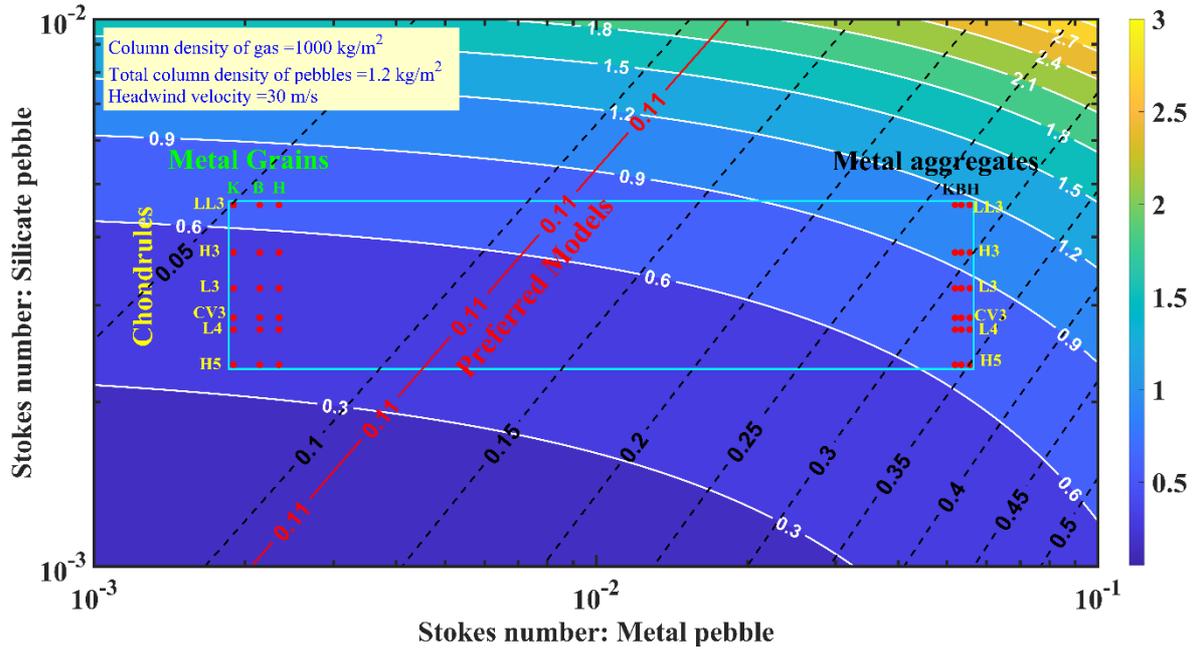

Figure 1: Final size of protoplanets after 2 My of Hill regime pebble accretion, for different Stokes numbers of silicate and metal pebbles. The plot shows protoplanet mass relative to Earth (white contours and shading) and protoplanet metal mass fraction (black contours) as functions of the Stokes numbers of metal and silicate pebbles. The red dots represent chondrule and metal grain Stokes numbers based on measured size and density in the Kelly (K), Bjurbole (B), and Hammond Down (H) chondrites, respectively. The turquoise box indicates the implied range of pebble Stokes numbers, including individual metal grains and metal grain aggregates. The red contour corresponds to an Earth-like (Fe+Ni)/Si mass ratio of 2.06 and defines our preferred models with protoplanetary masses between 0.35 and 0.7 Earth masses.

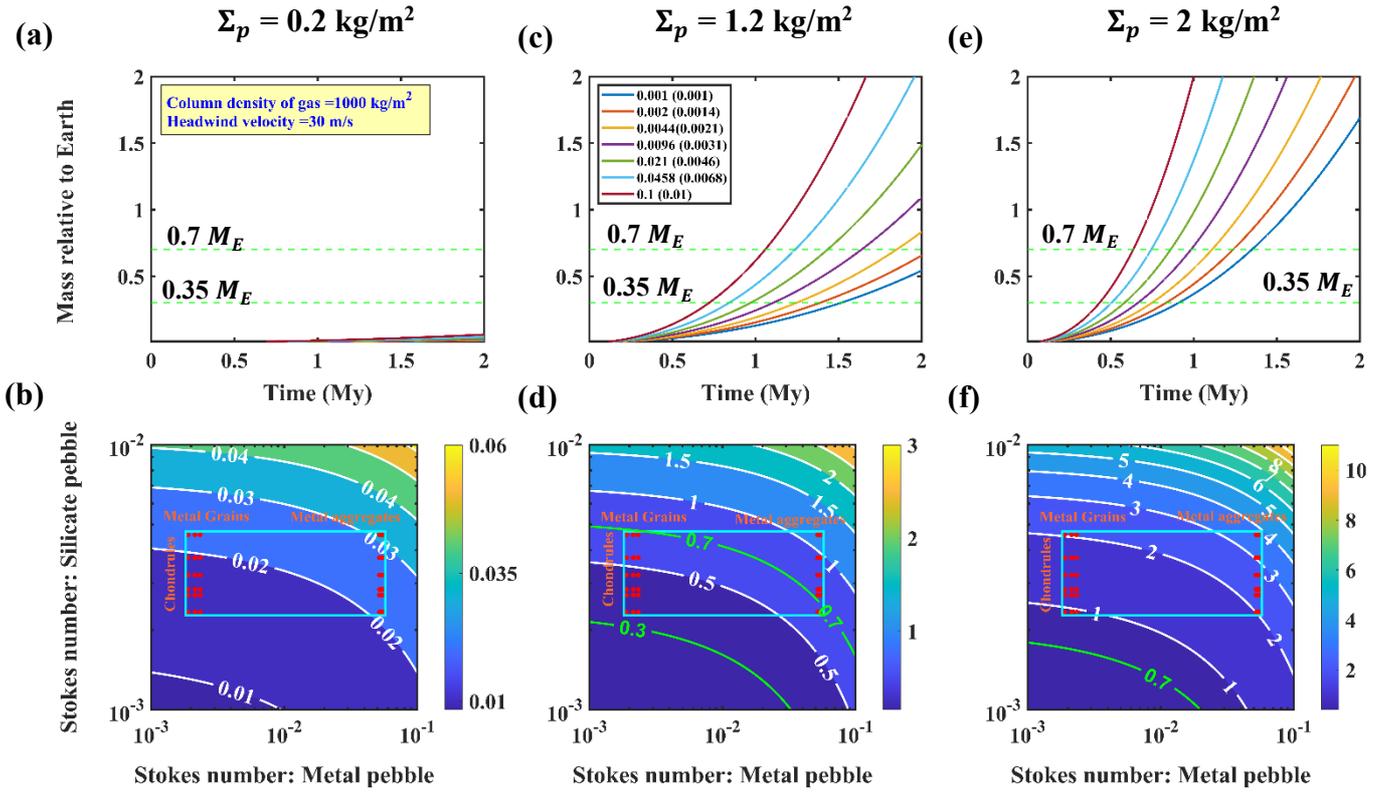

Figure 2: Protoplanet mass as a function of time in the Hill regime accretion. Top images (a, c, e): protoplanet mass (relative to Earth mass $M_E$) for various total pebble column densities $\Sigma_p$. The curves represent different combinations of Stokes numbers of metal pebbles (first number in legend in Figure 2c) and silicate pebbles (second number) at fixed gas density and headwind velocity. Dotted lines define a possible mass range for proto-Earth prior to the moon formation event. Bottom images (b, d, f): contours of protoplanet mass (relative to Earth) at 2 My for cases (a), (c), & (e) as functions of the Stokes numbers of metal and silicate pebbles. The red dots represent chondrule and metal grain Stokes number combinations based on measured size and density. The turquoise boxes represent the range of Stokes numbers including individual metal grains and metal aggregates.

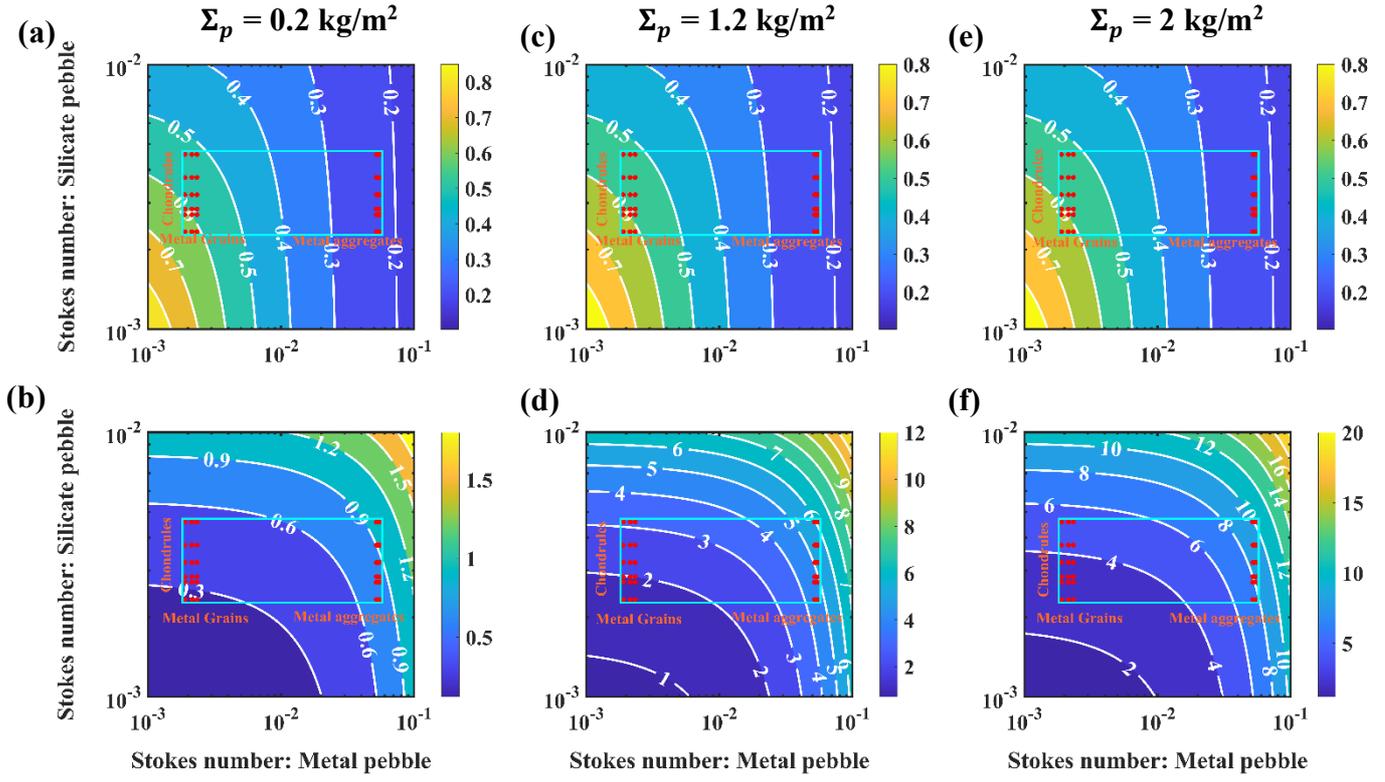

Figure 3: Top images (a, c, e): contours of pebble capture probabilities for a 0.7 $M_E$ protoplanet at 1 au as functions of the Stokes numbers of metal and silicate pebbles, for various pebble column densities $\Sigma_p$. Bottom images (b, d, f): contours of time-integrated mass flux (relative to Earth's mass) of drifting pebbles over 2 My in the Hill regime accretion at 1 au for cases (a), (c), & (e), as functions of the Stokes numbers for metal and silicate pebbles.

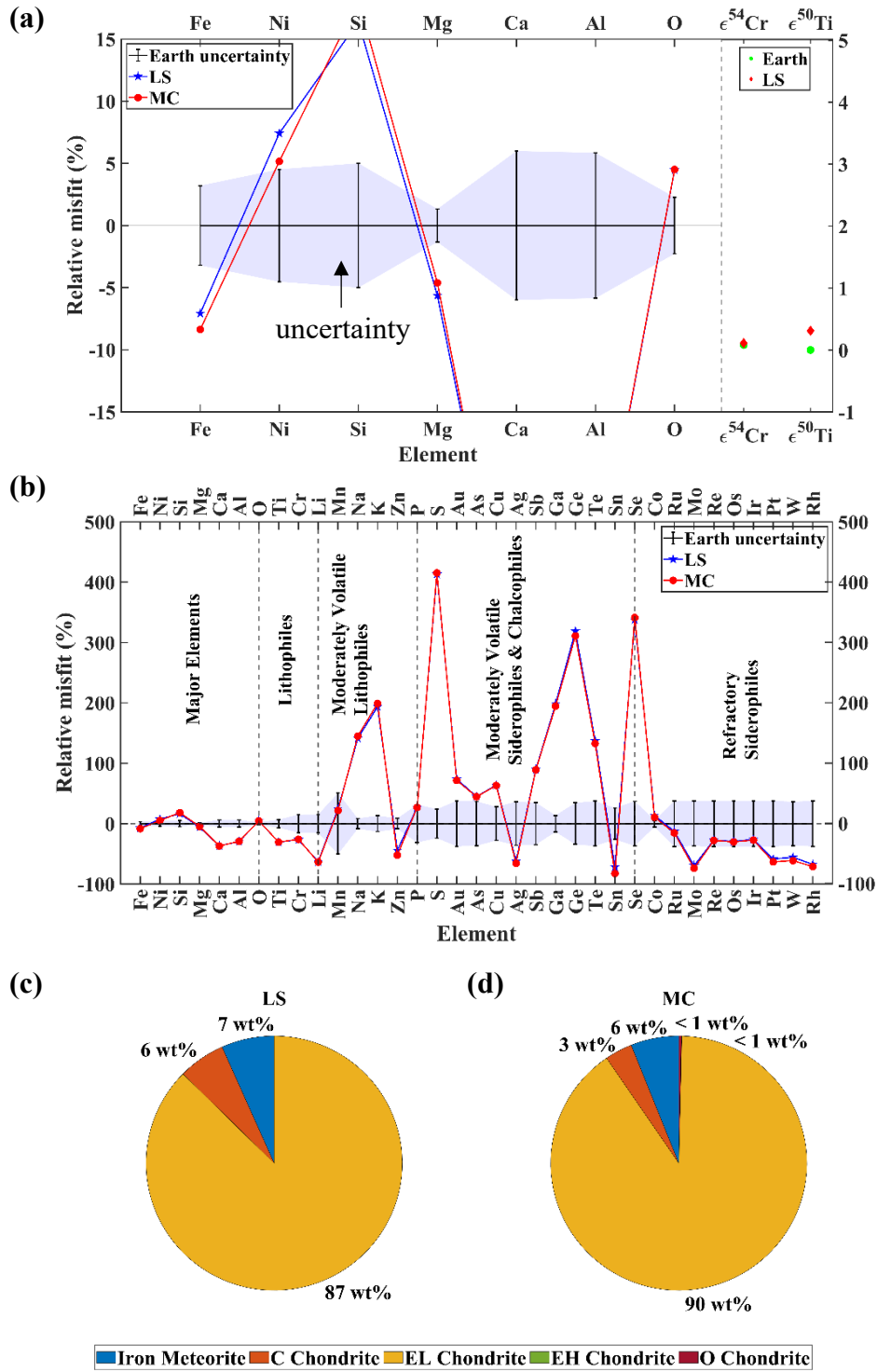

Figure 4: Chondrite model. (a) Relative misfits for major elements plus $\varepsilon^{54}$Cr and $\varepsilon^{50}$Ti values and (b) relative misfits for minor elements, with corresponding pebble proportions (c, d) obtained by Least-Squares (LS) and Markov-Chain Monte Carlo (MC) inversions. The blue shaded region in (a, b) is the elemental uncertainty for the bulk Earth from Wang et al. (2018).

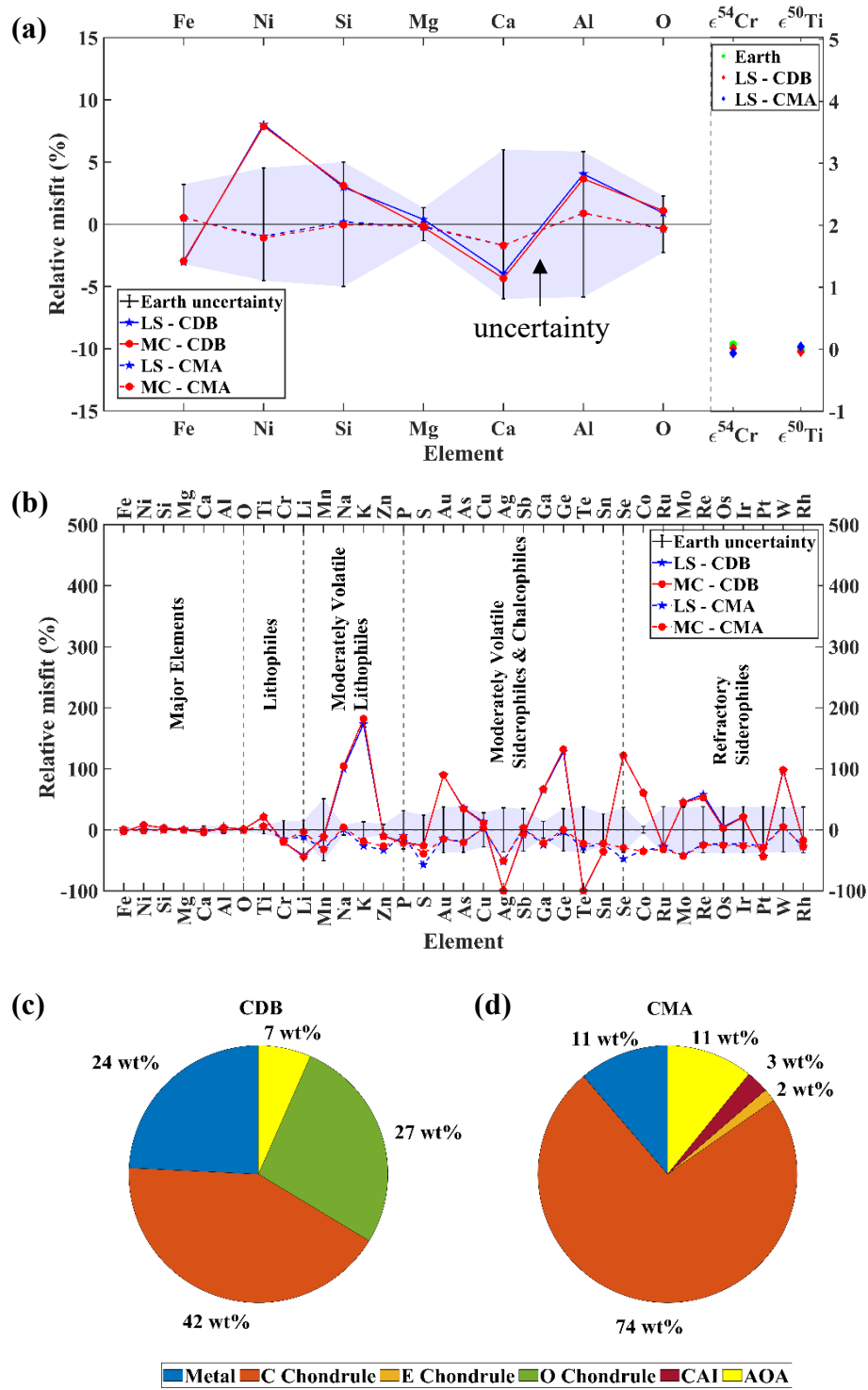

Figure 5: Component model. (a) Relative misfits for major elements plus $\varepsilon^{54}$Cr and $\varepsilon^{50}$Ti values and (b) relative misfits for minor elements as obtained by Least-Squares (LS) and Markov-Chain Monte Carlo (MC) inversions. The solid and dotted curves represent our CDB and CMA models, respectively. (c, d) Pebble proportions for both models obtained by LS inversion.

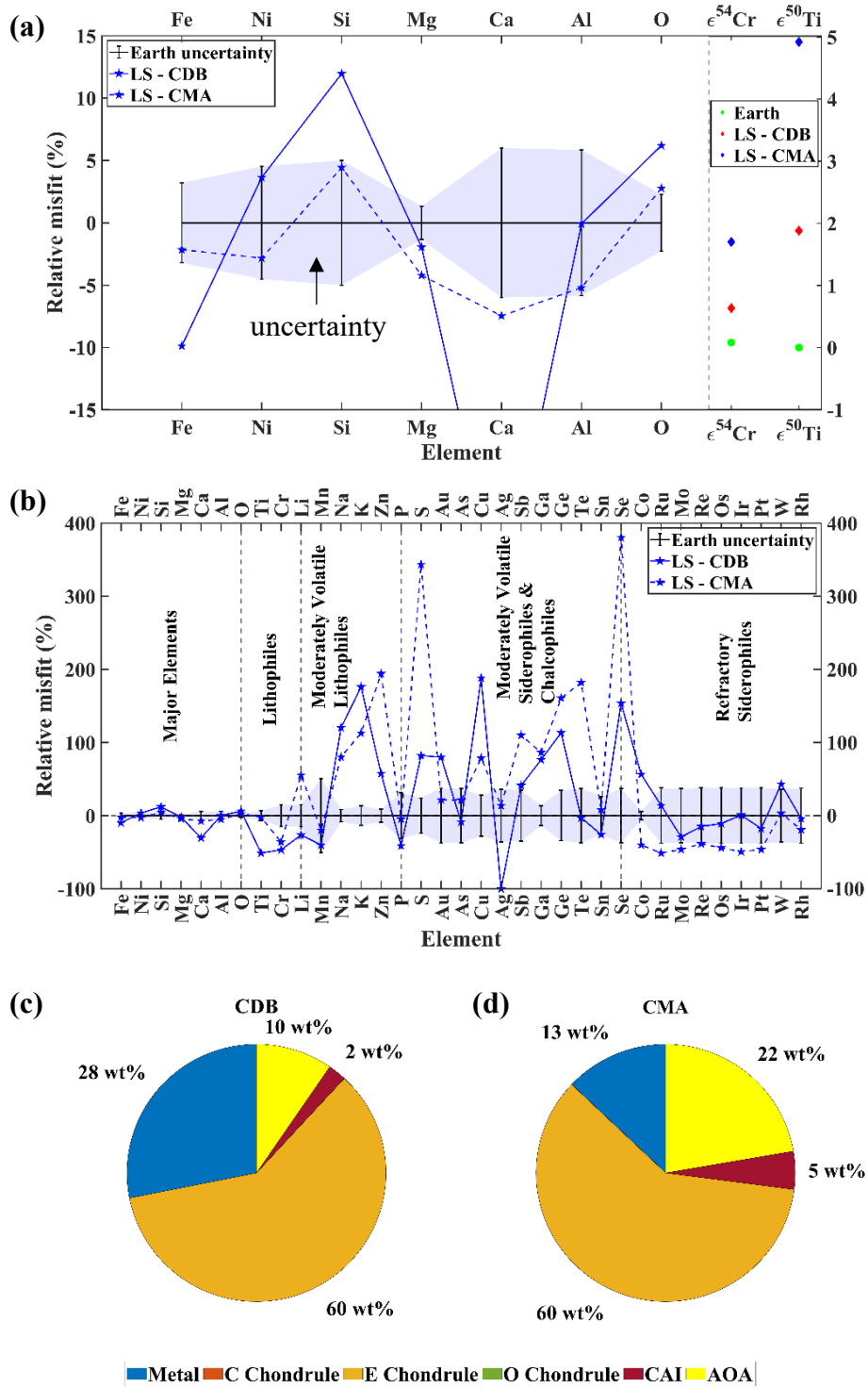

Figure 6: Constrained (E Chondrule = 60 wt%) Component model. (a) Relative misfits for major elements plus $\varepsilon^{54}$Cr and $\varepsilon^{50}$Ti values and (b) relative misfits for minor elements as obtained by Least-Squares (LS) inversion. The solid and dotted curves represent our CDB and CMA models, respectively. (c, d) Pebble proportions for both models obtained by LS inversion.

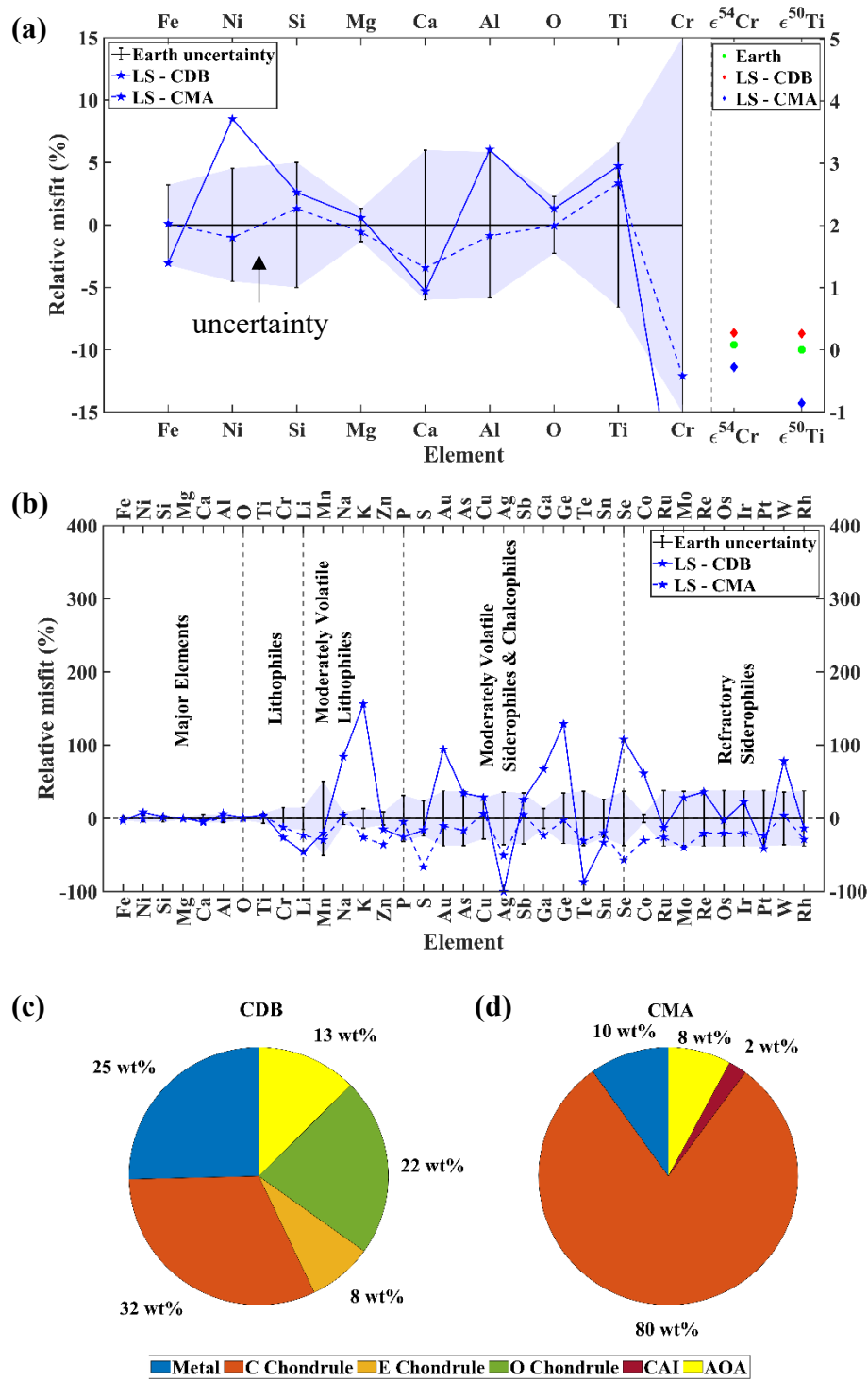

Figure 7: Component model with Ti and Cr included. (a) Relative misfits for major elements plus $\varepsilon^{54}$Cr and $\varepsilon^{50}$Ti values and (b) relative misfits for minor elements as obtained by Least-Squares (LS) inversion. The solid and dotted curves represent our CDB and CMA models, respectively. (c, d) Pebble proportions for both models obtained by LS inversion.

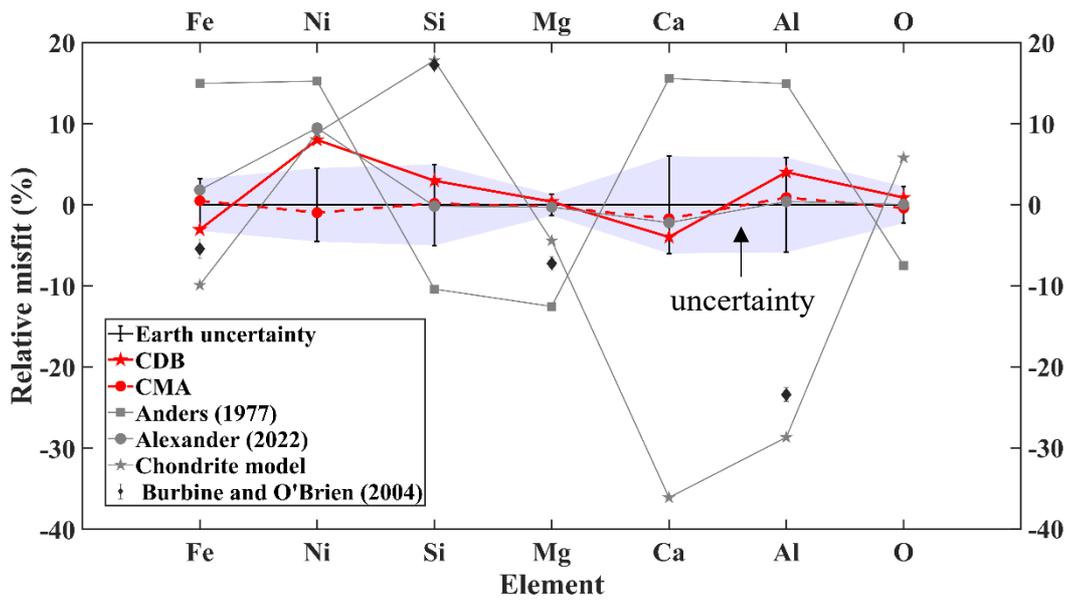

Figure 8: Major element comparison of four chondritic component models: CDB, CMA, Anders (1977), and Alexander (2022), versus our Chondrite model and the chondrite model of Burbine and O'Brien (2004).

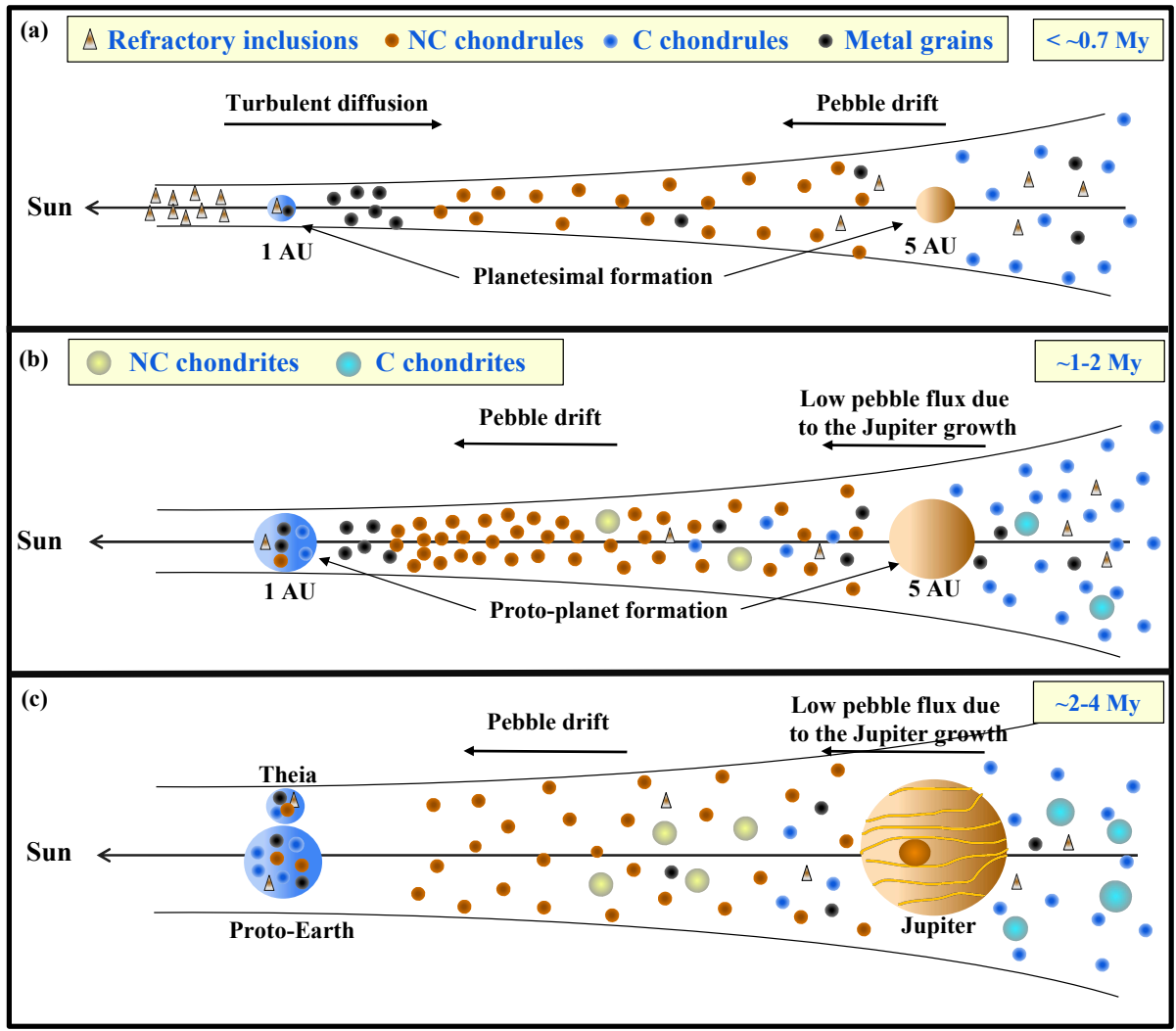

Figure 9: Schematic of the secular variation of the early solar system materials. (a) 0-0.7 My: The refractory grains condense close to the Sun. Some refractory grains scattered to the outer solar system by turbulent diffusion are ultimately incorporated in later-formed bodies. Metal grains and the early generation of chondrules form further away from the Sun. Planetesimal formation takes place at this stage by gravitational or streaming instabilities (Johansen and Lambrechts, 2017). (b) ~1-2 My: Chondrule production continues. Protoplanets form by pebble accretion. Chondrite accretion begins. (c) ~2-4 My: Chondrule production and pebble accretion continue with some addition of C chondrules to the inner solar system bodies due to incomplete capture by Jupiter (Johansen et al., 2021). Inner planets are not shown here.